\documentclass[referee]{raa}            % referee version: for submission

\usepackage{graphicx,times}             %for PS/EPS graphics inclusion, new
\usepackage{natbib}
\usepackage{amssymb,amsmath}
\bibpunct{(}{)}{;}{a}{}{,}
\graphicspath{{./}{Figures/}}
\usepackage{subcaption}
\usepackage{multirow}
\usepackage{listings}
\usepackage{xcolor}

\usepackage[pagebackref=true]{hyperref}

\begin{document}

  \title{All-sky Guide Star Catalog for CSST
\footnote{This research has been generously supported by the National Natural Science Foundation of China (Grant Nos. 12073047 and 12273077) and the National Key Research and Development (Grant No. 2022YFF0711500).}}
%   \subtitle{I. Place Your Subtitle Here}

   \volnopage{Vol.0 (20xx) No.0, 000--000}      %%preserved for Editor. DOn't remove!
   \setcounter{page}{1}          %%starting page, preserved for Editor. DOn't remove!

    \author{Hui-Mei Feng 
      \inst{1,2,3}
   \and Zi-Huang Cao
      \inst{3,4}
   \and Man I Lam 
      \inst{3}
   \and Ran Li
      \inst{3,2,4}
    \and Hao Tian
      \inst{3}
    \and Da-Yi Yin
      \inst{5}
    \and Yuan-Yu Yang
      \inst{5}
    \and Xin Zhang
      \inst{3}
    \and Dong-Wei Fan 
      \inst{3}
    \and Yi-Qiao Dong 
      \inst{3}
    \and Xin-Feng Li 
      \inst{6}
    \and Wei Wang
      \inst{7}
    \and Long Li
      \inst{7}
    \and Hugh R. A. Jones
      \inst{8}
    \and Yi-Han Tao
      \inst{3}
    \and Jia-Lu Nie
      \inst{3,4}
    \and Pei-Pei Wang
      \inst{3,4}
    \and Mao-Yuan Liu
      \inst{1}
    \and He-jun Yang
      \inst{9}
    \and Chao Liu
      \inst{3,4}
}

   \institute{Key Laboratory of Cosmic Rays (Tibet University), Ministry of Education, Lhasa 850000, China\\
        \and
            Institute for Frontiers in Astronomy and Astrophysics, Beijing Normal University, Beijing 102206, China
        \and
            National Astronomical Observatories, Chinese Academy of Sciences, Beijing 100012, China\\
        \and
            University of Chinese Academy of Sciences, Beijing 100049, China\\
        \and
            Shanghai Institute of Technical Physics, Chinese Academy of Sciences, Shanghai 200083, China\\
        \and
            Technology and Engineering Center for Space Utilization, Chinese Academy of Sciences, Beijing 100094, China\\
        \and
            Changchun Institute of Optics, Fine Mechanics and Physics, Chinese Academy of Sciences, Changchun 130033, China\\
        \and
            Centre for Astrophysics, University of Hertfordshire, Hatfield, UK\\
        \and 
            Beijing University of Technology, Beijing 100124, China\\
    {\it zhcao@nao.cas.cn}\\
    {\it mlam@nao.cas.cn}\\
    {\it liumaoyuan@163.com}\\
\vs\no
   {\small Received 2023 month day; accepted 202x month day}}

\abstract{ The China Space Station Telescope (CSST) is a two-meter space telescope with multiple back-end instruments. The Fine Guidance Sensor (FGS) is an essential subsystem of the CSST Precision Image Stability System (PISS) to ensure the required absolute pointing accuracy and line-of-sight stabilization. In this study, we construct the Main Guide Star Catalog for FGS. 
To accomplish this, we utilize the information about the FGS and object information from the $Gaia$ Data Release 3. We provide an FGS instrument magnitude and exclude variables, binaries, and high proper motion stars from the catalog to ensure uniform FGS guidance capabilities. 
Subsequently, we generate a HEALPix index, which provides a hierarchical tessellation of the celestial sphere, and employ the Voronoi algorithm to achieve a homogeneous distribution of stars across the catalog. 
This distribution ensures adequate coverage and sampling of the sky. The performance of the CSST guide star catalog was assessed by simulating the field of view of the FGS according to the CSST mock survey strategy catalog. 
The analysis of the results indicates that this catalog provides adequate coverage and accuracy. The catalog’s performance meets the FGS requirements, ensuring the functioning of the FGS and its guidance capabilities.
}

   \authorrunning{Feng et al. }            %author_head in even pages
   \titlerunning{All-Sky Guide Star Catalog For CSST}  % title_head in odd pages

   \maketitle
%% The author head (on even pages) and the title head (on odd pages) will be
%% automatically extracted from \author{} and \title{}. Whenever the title is too long,
%% you will be asked to supply a shorter one by inserting either \authorrunning{} or
%% \titlerunning{} before \maketitle. Anyway, you can specify your own heads.
%%
%%
%% Note: In the following text body of your manuscript, please note several differences from
%%       other major journals:
%% (1) \subsection{Please Capitalize the First Letter of Each Notional Word in Subsection Title}
%% (2) Please Capitalize the First Letter of Each Notional Word in all tables' captions

%
%________________________________________________ sections below
%

\section{Introduction} \label{sec:intro}

The Chinese Space Station Telescope (CSST), also called the China Space Survey Telescope, is a two-meter diameter space telescope that will be launched around 2025 to near-earth orbit. CSST will run independently at the same orbit as the Chinese Space Station, which benefits from on-orbit maintenance and upgrade (\citealt{2018An}). CSST is an off-axis three-reflector space telescope with a Fine Guidance Sensor (FGS) to provide a high-stability image system. The CSST is designed to achieve an 80\% energy concentration of PSF better than 0.15~arcsec, and a PSF ellipticity of better than 0.15. These figures of merit include the static and dynamic errors in the optical system and the back-end instruments.
The planned science goals of CSST cover a wide range of areas, including cosmology, the Milky Way and nearby galaxies, galaxies and active galactic nuclei (AGN), exoplanets and solar system objects, stellar science, astrometry, transients, and a wide variety of variable sources.

The Main Survey Camera (MSC) is the major instrument of CSST, which is expected to allocate about 70\% of observing time. CSST adapts the fixed filter or grating on each sensor and ensures each filter corresponds to a specific detector with the highest quantum efficiency within the detector's band. 30 Charge Coupled Devices (CCD) with fixed filters are deployed in MSC. Among them, 18 CCDs are allocated for multiband imaging with seven bands (NUV, $u$, $g$, $r$, $i$, $z$, $y$) and 12 for slitless spectrographs with three types of slitless spectrum grating (GU, GV, GI). Thus, the MSC can simultaneously observe 30 mosaic sky areas within a $1.1^{\circ}\times1.2^{\circ}$ Field of View (FOV) and can reach 26.3~mag at $G$ band in 300 seconds exposure time.
Besides, there are also other instruments mounted on the CSST, including the Integral Field Spectrograph (IFS), the Multi-Channel Imager (MCI), the Cool Planets Imaging Coronagraph (CPI-C), and the High Sensitivity Terahertz Detection Module (HSTDM). IFS, MCI, CPI-C, and FGS deploy on the Shared Focal Plane (SFP) with FGS, and HSTDM deploys on an independent focal plane without FGS. Each instrument will be allocated about 5\% observation time in the total telescope lifetime in orbit. 
 
CSST is designed to achieve the performance that the PSF's 80\% energy concentration does not exceed 0.15~arcsec, and the PSF ellipticity remains below 0.15. However, the CSST absolute pointing accuracy and image stabilization accuracy are designed to achieve no more than 10~arcsec. Meanwhile, with the assistance of a guide star, these accuracies can be significantly improved to 5~arcsec and 0.05~arcsec, respectively (\citealt{Zhan2021}). To accomplish this, FGS plays a critical role in achieving these improved accuracies and ensuring the necessary quality factors for scientific objectives. Drawing from the experience of previous space telescope missions, the guide star catalog assumes great importance in the functioning of the FGS.
In this study, we present the details of the CSST guide star catalog (CSST-GSC) solution and conduct a preliminary test and analysis of the guide star catalog's performance based on the theoretical technical performance of the FGS and the current simulated sky survey planning.

The paper is organized as follows. Section~\ref{sect:fgs_requirements} introduces the FGS characteristics and requirements for the guide star catalogs. Section~\ref{sect:input_catalog} introduces the input catalogs for catalog building and the mock survey strategy catalog for verification. Section~\ref{sect:Method} presents the FGS guide star catalog building and validation methods. Section~\ref{sect:Res} shows the results related to input catalogs preprocessing, catalog homogenization, and our catalog performance. Section~\ref{sect:mgsc_release} delineates the catalog release. Finally, Section~\ref{sect:summary} summarizes this study.

\section{Requirements To Guide Star Catalog} \label{sect:fgs_requirements}

%------------------------------------------------------------
   \begin{figure}
   \centering
   \includegraphics[width=1.0\textwidth, angle=0]{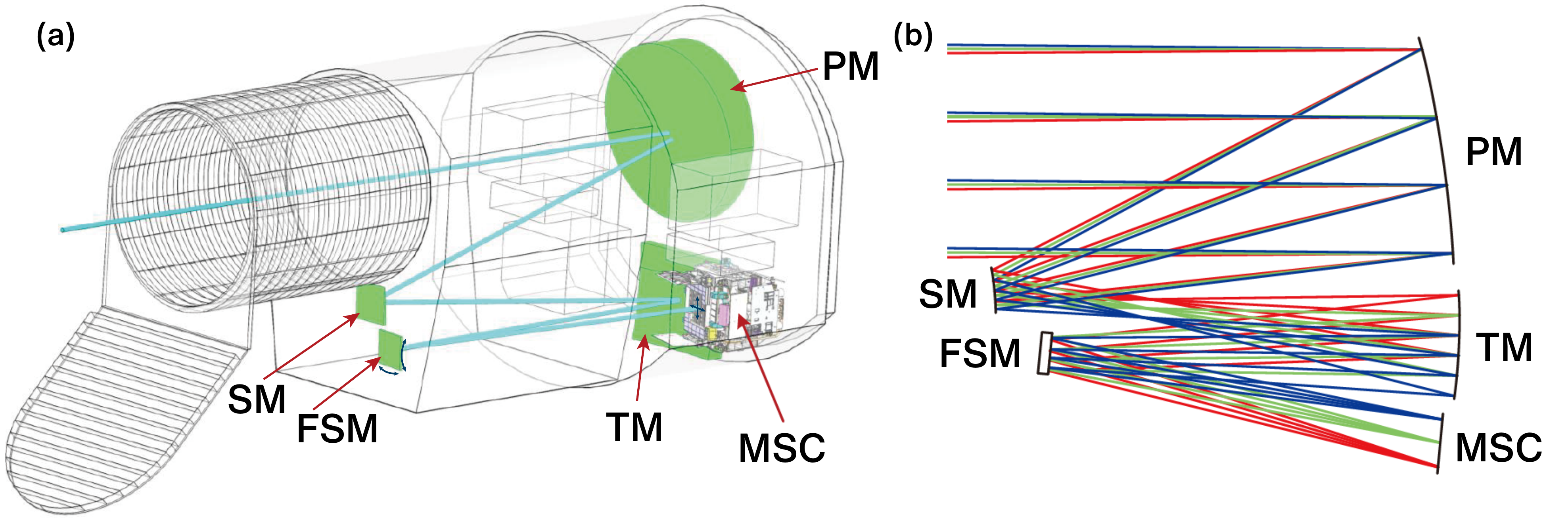}
   \caption{A diagram of the light path of CSST off-axis three-reflector to illustrate FGS guiding (modified from \cite{Zhan2021}). In sub-figure (a), green parts represent the mirrors, including the primary mirror (PM), the secondary mirror (SM), the tertiary mirror (TM), and the Fast Steering Mirror (FSM). In sub-figure (b), different light beam colors represent the different wavelengths, and the FSM reflects the light to the MSC for observation with FGS guiding. We show the FSM movement direction and the corresponding light beam movement direction on MSC in blue two-way arrows when the FGS is guiding. }
   \label{Sec2_Fig001}
   \end{figure}
   
%------------------------------------------------------------
   \begin{figure}
   \centering
   \includegraphics[width=1.0\textwidth, angle=0]{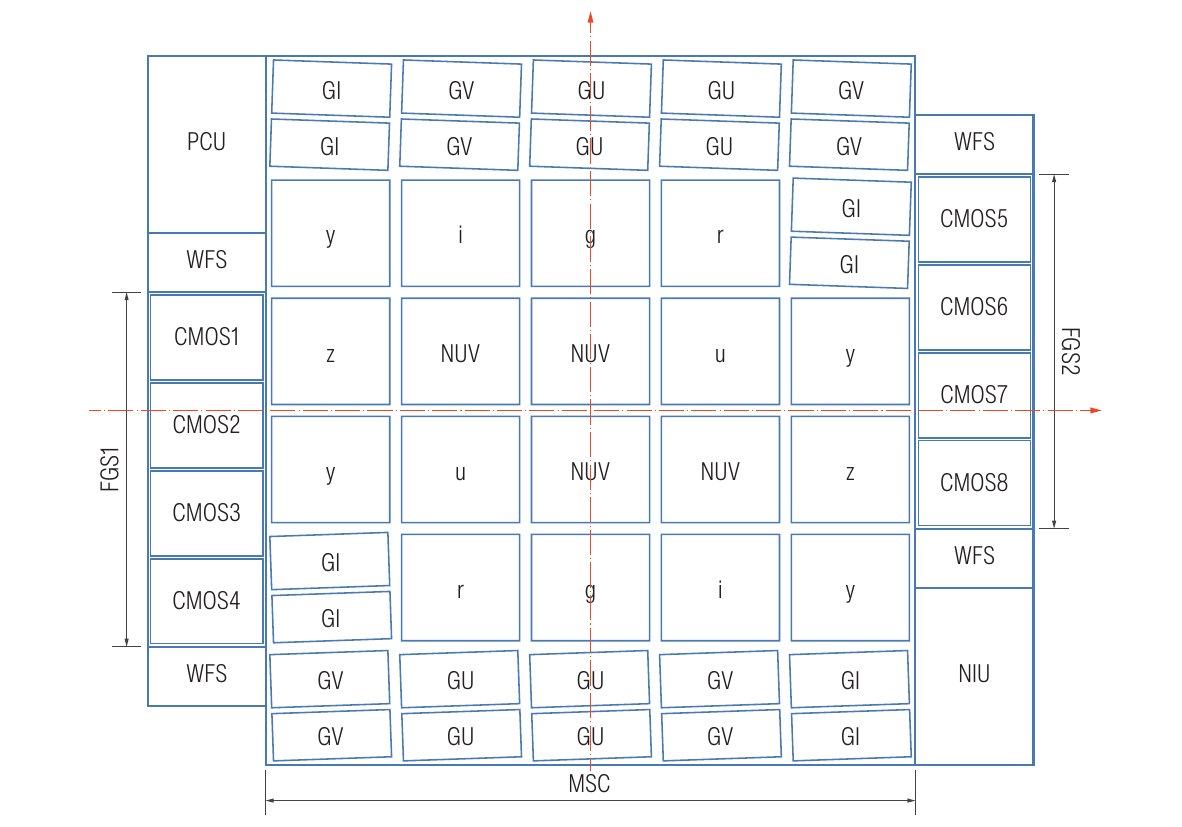}
   \caption{The Main Survey Camera (MSC) focal plane distribution of CSST. On the focal plane, a total of 30 CCDs are deployed in MSC, 18 for multiband imaging with seven bands (NUV, $u$, $g$, $r$, $i$, $z$, $y$) 
 and 12 for slitless spectrographs with three types of slitless spectrum grating (GU, GV, GI). The layout of these CCDs is symmetrically arranged along the center to facilitate observation. The focal panel also includes one Photometric Calibration Unit (PCU), one Near-infrared Unit (NIU), four Wavefront Sensors (WFS), and two Fine Guidance Sensors (FGS) (\citealt{Zhan2021}). FGS1 is located on the left side and consists of four CMOS detectors (CMOS1, CMOS2, CMOS3, CMOS4), where CMOS1 and CMOS2 serve as primary sensors, and CMOS3 and CMOS4 serve as backup sensors. FGS2 is situated on the right side and comprises four CMOS detectors (CMOS5, CMOS6, CMOS7, CMOS8), where CMOS1 and CMOS2 serve as primary sensors, and CMOS7 and CMOS8 serve as backup sensors. We establish a cartesian coordinate system on the MSC, with the forward directions defined as right and up, and the MSC center designated as the origin point.
   }
   \label{Sec2_Fig002}
   \end{figure}

%------------------------------------------------------------
   \begin{figure}
   \centering
   \includegraphics[width=0.9\textwidth, angle=0]{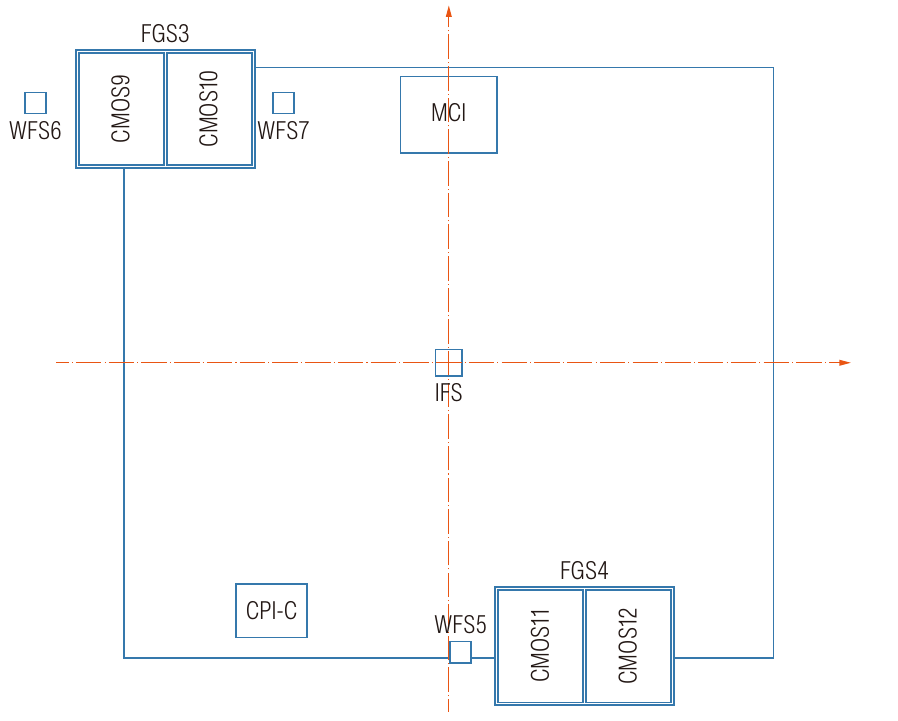}
   \caption{The Shared Focal Plane (SFP) distribution of CSST. The SFP center is the original point to build the coordinate. Two FGS are deployed on the SFP. FGS3 is on the upper left, and FGS4 is on the lower right. Each FGS contains two CMOS detectors (FGS3: CMOS9, CMOS10 and FGS4: CMOS11, CMOS12) without a backup sensor. We establish a cartesian coordinate system on the SFP, with the forward directions defined as right and up, and the SFP center designated as the origin point.}
   \label{Sec2_Fig003}
   \end{figure}

%------------------------------------------------------------
   \begin{figure}
   \centering
   \includegraphics[width=1.0\textwidth, angle=0]{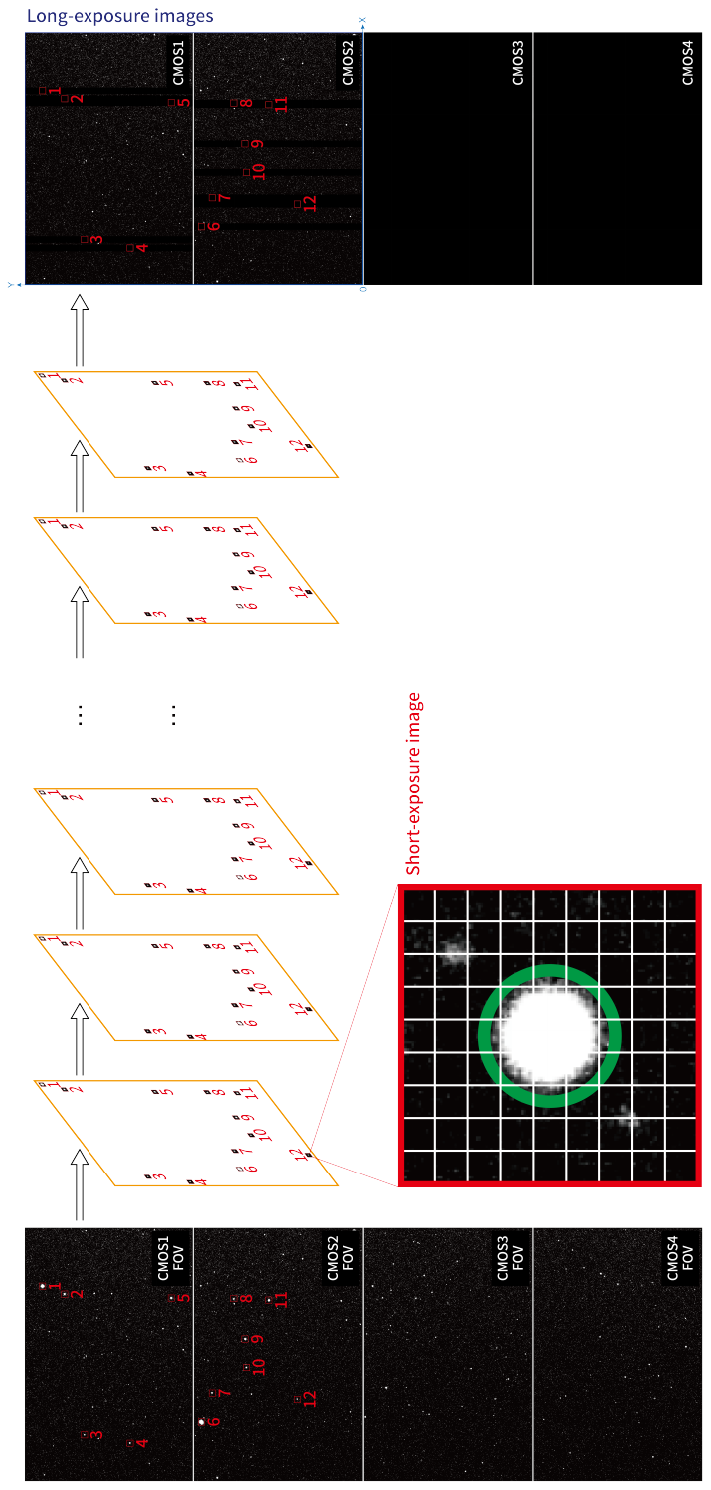}
   \caption{The simulated FGS output images in the long (left-hand images) and short-exposure (right-hand images). In this figure, we take the FGS1, for example, including the activated primary sensors (CMOS1, CMOS2) and the inactivated backup sensors (CMOS3, CMOS4). The images labeled CMOS1-4 on the left represent the sky areas in the FGS1 FOV. The yellow frames in the upper middle represent many sets of ROIs captured by the CMOS1 and CMOS2 in high frequency in short-exposure mode. The tiny red squares in yellow frames represent the ROI windows, labeled with serial numbers in red characters. The red frame in the lower middle is a short-exposure image example. It is a zoom-in of one of the ROI windows, comprising an ROI of 9$\times$9 pixels. The FGS selected the star in this example as a guide star, marked as a green circle. The images labeled CMOS1 and CMOS2 on the right represent the images captured by FGS1 in long-exposure mode, and the empty readout regions in black due to the ROI on their rows. The CMOS3 and CMOS4 images are black, representing CMOS3 and CMOS4 have no outputs. The row and column directions of the image follow the blue arrows, marked as X and Y.}
   \label{Sec2_Fig004}
   \end{figure}

%------------------------------------------------------------
   \begin{figure}
   \centering
   \includegraphics[width=1.0\textwidth, angle=0]{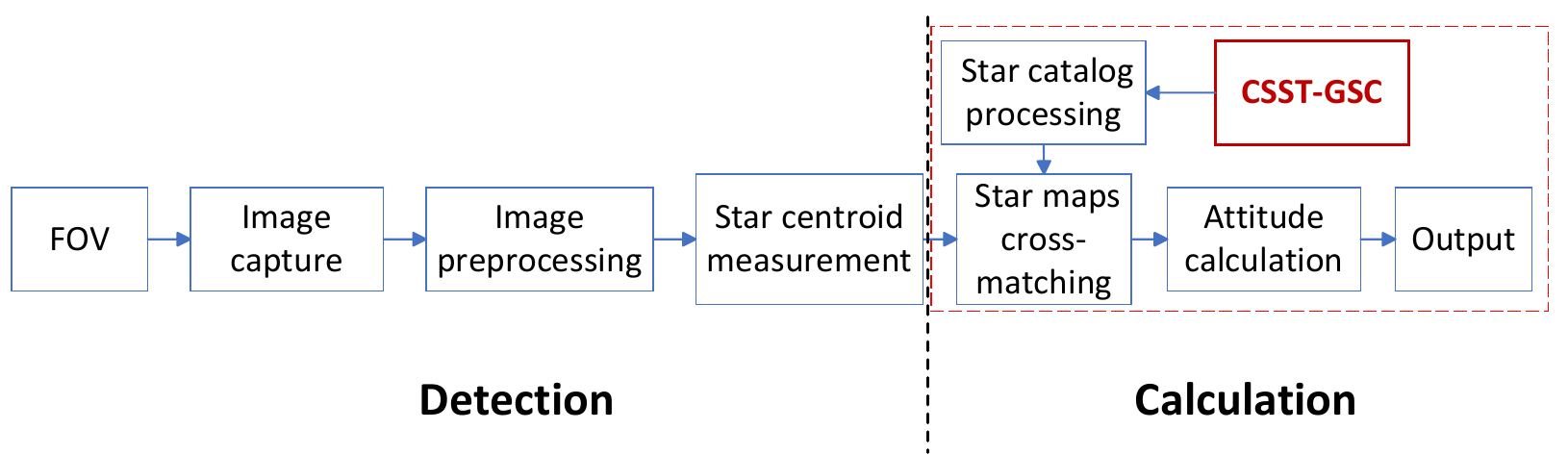}
   \caption{The workflow for FGS involves several steps. Image processing tasks include denoising, background estimation, and threshold segmentation. The star celestial coordinates (R.A., Dec.) from the CSST-GSC catalog should be converted to the FGS measurement coordinates (X, Y) for star maps cross-matching. Finally, the FGS generates output control signals for PISS. }
   \label{Sec2_fgs_cal}
   \end{figure}

%_______________________________________________________________________________
\begin{table}
\centering
\begin{minipage}[]{\textwidth}
\caption{The expected FGS performance and configuration parameters.
\label{Sec2_Tab01}} \end{minipage} \\
\setlength{\tabcolsep}{3mm}
\small
 \begin{tabular}{ccc}
  \hline
  \hline\noalign{\smallskip}
 No. & Items & Contents\\
  \hline\noalign{\smallskip}
 1  &  MSC Primary Detectors  &  CMOS1, CMOS2 (FGS1)/ CMOS5, CMOS6 (FGS2)  \\
 2  &  MSC Backup Detectors  &  CMOS3, CMOS4 (FGS1) / CMOS7, CMOS8 (FGS2) \\
 3  &  SFP Primary Detectors  &  CMOS9, CMOS10 (FGS3) / CMOS11, CMOS12 (FGS4) \\
 \multirow{2}*{4} &  \multirow{2}*{MSC Effective Area}  &  $\ge$0.12 $deg^{2}$ (Primary+Backup Detectors)  \\
 ~ &  ~  &  $\ge$0.06 $deg^{2}$ (Primary or Backup Detectors) \\
 5  &  SFP Effective Area  &  $\ge$0.06 $deg^{2}$  \\
 6  &  CMOS Pixel Size  &  7.5 $\mu$m  \\
 7  &  CMOS Pixel Array  &  7680 $\times$ 11264 pixels \\
 8  &  CMOS Pixel Scale  &  0.0375 arcsec/$\mu$m  \\
 9  &  CMOS QE  &  Refer to  Figure~\ref{Sec3_Fig001} \\
 10  &  Optical Transmission  &  Refer to  Figure~\ref{Sec3_Fig001} \\
 11  &  System Transmission  &  Refer to  Figure~\ref{Sec3_Fig001} \\
 12  &  Frame Frequency  &  $\ge$100 fps $@$ (256 rows, 14 bit)  \\
 13  &  ROI Size  &  5$\times$5, 7$\times$7, 9$\times$9, or 15$\times$15 pixels\\
 14  &  Size For Transfer   &  15$\times$15 pixels\\
 15  &  Maximum ROI Number   &  16\\
  \noalign{\smallskip}\hline
  \hline
\end{tabular}
\end{table}

\subsection{The Fine Guidance Sensor} \label{sect:fgs_features}
The Precision Image Stability System (PISS) is used to correct the observation or pointing direction in a limited range and stabilize the image in CSST observation. FGS and the Fast Steering Mirror (FSM) are two essential components of the PISS. The FSM, known as the "three-in-one" mirror, possesses multiple functionalities, including focal length adjustment, light path switching between MSC, SFP, and HSTDM, as well as high-frequency sheering for guiding purposes. The specific placement of the FSM within the light path is illustrated in Figure~\ref{Sec2_Fig001}. By utilizing high-frequency stellar position sampling and calibration data from the FGS, the PISS can dynamically drive the FSM in closed-loop control, enabling high-precision and real-time adjustments of the light axis. This capability effectively corrects for spacecraft's pointing errors and reduces light-axis vibrations.

FGS adopts a large-frame custom-made Complementary Metal Oxide Semiconductor (CMOS) detector and aerospace-grade FPGA to provide accurate guide stars' position, their derived results, and control signals with high efficiency and low energy consumption.
Four FGSs are prepared and installed in CSST, of which two FGSs are located on the MSC. As shown in Figure~\ref{Sec2_Fig002}, FGS1 and FGS2 are deployed on both sides of the MSC, and each contains four CMOS detectors, including two primary and two backup detectors. The other two FGSs are deployed on the SFP. As shown in Figure~\ref{Sec2_Fig003}, FGS3 and FGS4 are on the upper left corner and lower right of the SFP, each containing two CMOS detectors without a backup sensor. The deployment of FGSs in close proximity to the field of view (FOV) is crucial for achieving high measurement precision. Additionally, to maximize the astrometric baseline between two FGSs, their layout in the MSC and SFP is illustrated in Figure~\ref{Sec2_Fig002} and Figure~\ref{Sec2_Fig003}, respectively. Detailed parameters of the FGS are provided in Table~\ref{Sec2_Tab01}.

Backup sensors do not work on regular observation. Suppose the primary sensors can not work properly or there are insufficient guide stars in the primary sensors’ FOV or FGS in calibration observation. In these cases, we can activate the backup sensors and make them work with or aside from the primary sensors. The "activate" procedure includes powering on, self-checking, cooling down the sensors, capturing the first image, accessing the PISS system, and more. The "work" mentioned here includes simultaneously capturing the images in FOV in short- and long-exposure modes.

Operation and observation simultaneously in both short- and long-exposure modes is a feature of the CSST FGS. In the short-exposure mode, the FGS captures high-frequency samples of guide stars within a predefined ROI. Conversely, the long-exposure mode covers the entire FGS sensor area, excluding the ROI-occupied rows. Based on the advantages of this design, each FGS can provide an extra FOV of 0.06 square degrees for primary sensors and 0.12 square degrees for all sensors. In the long-exposure mode, the FGS exposure time can be extended to match the maximum exposure time of MSC, allowing for photometric image analysis in combination with scientific observation data. Figure~\ref{Sec2_Fig004} illustrates the simulation of FGS output images and their relationship in the long-exposure and short-exposure modes. This feature makes the FGS can be used as an auxiliary observation instrument. The working process of the FGS consists of two main parts: 'Detection' and 'Calculation', as depicted in Figure~\ref{Sec2_fgs_cal}. This catalog is utilized as an input in this process, providing theoretical celestial coordinates of stars for the FGS.

FGS is crucial for space telescopes to achieve sub-arcsecond or higher accuracy measurements. Many space observation platforms, launched or under development, have incorporated this instrument for high-precision pointing. These platforms include the Hubble Space Telescope (HST), the Spitzer Space Telescope (SST), the James Webb Space Telescope (JWST), the Euclid Space Telescope, and CSST. The success of HST made astronomers and engineers realize the importance of FGS. The $Hubble$ $Space$ $Telescope$ ($HST$) have been equipped with 3 interferometric FGS to guarantee the pointing stability with 7 milliseconds of arc (\citealt{1988JGCD...11..119B}) and the astrometric precision with 0.3 arcsec (\citealt{2008IAUS..248...23B}). There are many scientific studies yielded by $HST$ FGS, including accurate parallaxes of astrophysical interesting stars and mass estimates for stellar companions obtained through FGS interferometric astrometry. (\citealt{2017Astrometry}), acquiring high angular resolution survey of massive OB stars in the Cygnus OB2 (\citealt{2014AJ....147...40C}), measuring precise diameters and shapes of asteroids even suspected to be binary bodies (\citealt{2002ESASP.500..517T}). Furthermore, the FGS can also contribute to the calibration of engineering parameters of the spacecraft in orbit. For instance, it can be used to correct installation errors of star sensors (\citealt{2021JPhCS1971a2015G}).

\subsection{FGS Constraints On Catalog} \label{sect:constraints}
FGS imposes specific requirements on the guide star catalog due to its operation in the short-exposure mode. In this mode, the FGS captures high-frequency samples of guide stars within a predefined ROI, allowing for rapid readout, star image calculation, and feedback to the Precision Image Stability System (PISS) for precise control. Considering the CSST Point Spread Function (PSF) with an 80\% energy concentration size, a guide star occupies approximately 4$\times$4 pixels on the FGS sensor. The ROI size can be configured within the range of 5$\times$5 to 15$\times$15 pixels. Therefore, guide stars should be isolated within a 5$\times$5 pixel area, with an approximate separation of 0.6 arcsecs between two stars. This requirement guarantees that closely located guide stars are adequately resolved and independent within the ROI. In the current design, the exposure duration for FGS short-exposure modes is expected to be less than 10 milliseconds. This ensures that the frequency of FGS guidance is not less than 100 Hz. The corresponding guide star magnitudes range from 8 to 15 magnitudes in the $Gaia$ $G$ band.

Under standard operating conditions, each FGS can open up to 16 ROIs based on the positions of target guide stars. All guide stars' ROI images, with a fixed size of 15$\times$15 pixels, would be transferred to the ground with other CSST working parameters. A selection process is employed to iteratively choose the best 9 guide stars from the available ROIs to ensure optimal high-frequency guidance. This selection is based on the guide star's brightness, position, and shape. On the other hand, as described in \cite{Bosco2015}, it requires a minimum of three targets per detector (FOV = 0.1 $\times$ 0.1 deg) to determine the sky position in theory. Hence, the density of guide stars within a limited magnitude range should be taken into account in the catalog, considering the FOV of the FGS. 

In addition, the density of available guide stars could be reduced in specific areas, such as high Galactic latitude regions. Activating backup sensors would be a suitable approach to ensure FGS performance in these areas. However, it's essential to consider that activating the backup sensors has drawbacks, including increased power consumption, a higher risk of FGS damage, and reduced reliability during its whole in-orbit life. Furthermore, the sensors require a prolonged cool-down period before being used effectively. Therefore, it is advisable to maintain continuous operation of the sensors, rather than frequently activating and deactivating them, to reduce the overhead and mitigate these risks. Lastly, it is crucial to carefully plan and utilize the FGS system's limited storage and computing resources. Efficient resource management is crucial to ensure optimal functionality and performance.

\subsection{CSST Guide Star Catalog (CSST-GSC)} \label{sect:fgs_catalog}
Three types of guide star catalogs are requested in the CSST sky survey:

\begin{itemize}
\item [1]
The CSST Main Guide Star Catalog (CSST-MGSC) is an essential resource that provides comprehensive coverage of the sky and encompasses potential objects that can be utilized as guide stars for FGS. The catalog should incorporate high-precision astrometric parameters based on the International Celestial Reference System (ICRS). Besides, the catalog needs to offer reliable photometric parameters close to the unfiltered measurements of FGS. 
The catalog will be maintained and upgraded on the ground. The CSST-MGSC plays different roles in different stages of the CSST mission. It serves as a fundamental tool for verifying the effectiveness of the sky survey strategy, facilitating the simulation of guide star images, and enabling research on sky survey strategy algorithms, when CSST is in R\&D and manufacturing. Furthermore, it acts as the input for generating both the Uploaded Guide Star Catalog and the On-board Guide Star Catalog when the CSST is in orbit.
\item [2]
The Uploaded Guide Star Catalog (CSST-UGSC) is employed for constellation matching during the CSST pointing stage and for tracking specific guide stars at the CSST exposing stage. The CSST-UGSC is built upon the CSST-MGSC, and the conversion from CSST-MGSC to CSST-UGSC requires careful consideration of various instrumental effects of astrometric precision and FGS parameters, as well as calibration parameters resulting from laboratory tests and on-orbit experiments. This conversion involves selecting appropriate guide stars for each FGS sensor for every CSST observation and converting their coordinates to FGS measurement coordinates while accounting for static aberrations of CSST optics, physical construction, dynamic distortion of active optics, and astrometric effects on orbit. The CSST-UGSC should be compiled and uploaded to the CSST approximately one week before observation, and enables efficient and reliable search and lock-on of pre-selected guide stars by the FGS, reducing search time and improving observation efficiency. 
\item [3]
The On-board Guide Star Catalog (CSST-OGSC) is a limited-sized catalog stored on board. It is derived from the CSST-MGSC and follows the same processing methodology as the CSST-UGSC. The CSST-OGSC is designed to cover a specific sky area for an extended observation period under the sky survey strategy. The catalog file is stored in a compression form in the FGS. The CSST-OGSC is prepared for calibration and testing in orbit and serves as an alternative resource to the CSST-UGSC in case of unforeseen issues with the uploaded catalog.
\end{itemize}

In summary, the CSST-MGSC, CSST-UGSC, and CSST-OGSC are interconnected components of the CSST guide star catalog. Each catalog serves a specific role within the system. The CSST-MGSC provides highly accurate astrometric and photometric parameters for celestial objects and is the foundation for the CSST-UGSC and CSST-OGSC. As a result, the precision and performance of the CSST-MGSC directly impact the capabilities of the FGS and play a crucial role in all stages of the CSST mission. Therefore, the primary objective of this study is to build a primary version of CSST-MGSC and ensure it satisfies the FGS requirements.

\section{Data} \label{sect:input_catalog}
The input star catalog refers to the raw data used to construct the guide star catalog. Therefore, the quality and accuracy of the input star catalog directly affect the performance of the guide star catalog. The input star catalog features we focus on include the sky coverage, depth, and accuracy of the astrometry parameters, including position and proper motion. Besides, the most recent version of the catalog is also essential, which represents the new data is added or processing techniques are improved. The $Gaia$ mission, based on the principles of the European Space Agency’s (ESA) $Hipparcos$ mission and launched on 19 December 2013 \citep{2016Gaia}, is the latest data released of $Gaia$ Data Release 3 ($Gaia$ DR3) \footnote{https://gea.esac.esa.int/archive/} and has been released on 13 June 2022 \citep{Vallenari2022}. It is currently one of the most precise and uniform full-sky catalogs available. Accordingly, we used $Gaia$ DR3 as the input for our guide star catalog. Besides, we adapted a mock survey strategy catalog to verify our guide star catalog that followed the survey planning strategy still in development. The mock survey strategy catalog recorded all CSST observation center locations in celestial coordinates. Using this catalog, we could project the guide stars from CSST-MGSC onto the FGS measurement coordinate system to determine if the number of guide stars was sufficient for FGS or the effect on the FGS performance.

\subsection{Input Catalogs} \label{subsect:data_dr3}
We select $Gaia$ DR3 main source catalog and its additional catalogs as the input catalogs. The $Gaia$ DR3 main source catalog includes around 1.81 billion sources, including 6.65 million (0.37\%) QSOs candidates and 4.84 million (0.27\%) galaxy candidates. 
Accurate $G$ magnitudes are provided for around 1.806 billion sources (99.7\%). Besides, BP and RP magnitudes are also provided for about 1.54 billion (85.11\%) and 1.55 billion (85.83\%) sources, respectively.
Around 1.46 billion sources are covering all sky with full astrometric solution, including positions, parallaxes, proper motions, and quality parameters like Renormalized Unit Weight Error (RUWE), in a magnitude range from 3 to 21 at the $G$ band, among them about 585 million (32.31\%) sources with 5-parameter, 882 million (48.7\%) sources with 6-parameter \citep{Vallenari2022}.  

We are concerned about the accuracy and precision of $Gaia$ DR3 measurements relying on various factors. The actual uncertainties may vary depending on the specific properties of the sources of interest. Fortunately, $Gaia$ DR3 provides highly accurate astrometric and photometric measurements for sources within the required magnitude range. The positional uncertainty typically ranges from 0.01 to 0.02 mas for five-parameter astrometry and 0.02 to 0.03 mas for six-parameter astrometry. The proper motion uncertainty is generally between 0.02 and 0.03 mas/yr for five-parameter astrometry and 0.02 and 0.04 mas/yr for six-parameter astrometry. The mean $G$-band photometry uncertainty falls within the 0.3 to 1.0 millimagnitude(\citealt{brown2021gaia}). 

Besides the main source catalog, $Gaia$ provides rich data products that benefit our study. Over 14 million pieces of time-domain information are processed and published, including variable sources, galaxies, and sources in the $Gaia$ Andromeda Photometric Survey (GAPS) \citep{Vallenari2022}. Further explorations of variables are also released at the same time, such as processing and validation of Cepheid and RR Lyrae \citep{Ripepi2022}, analysis of Long-Period Variable (LPV) candidates \citep{Lebzelter2022} and ellipsoidal variables \citep{Gomel2022}, analyzing the patterns of magnitude-color variations for solar-like variables \citep{Distefano2022}, classification of the variable young stellar object candidates \citep{Marton2022}. In addition, About 800,000 non-single stars (i.e., binary systems) are included in the $Gaia$ DR3 variable catalog \citep{Vallenari2022}, which include astrometric binaries, spectroscopic binaries, and eclipsing binaries. The variable catalog and the binary star catalog \citep{Halbwachs2022, Holl2022} are adopted to reject the corresponding sources in this study.

%------------------------------------------------------------
\begin{figure}
  \begin{minipage}[t]{0.5\linewidth}
    \centering
    \includegraphics[scale=0.43]{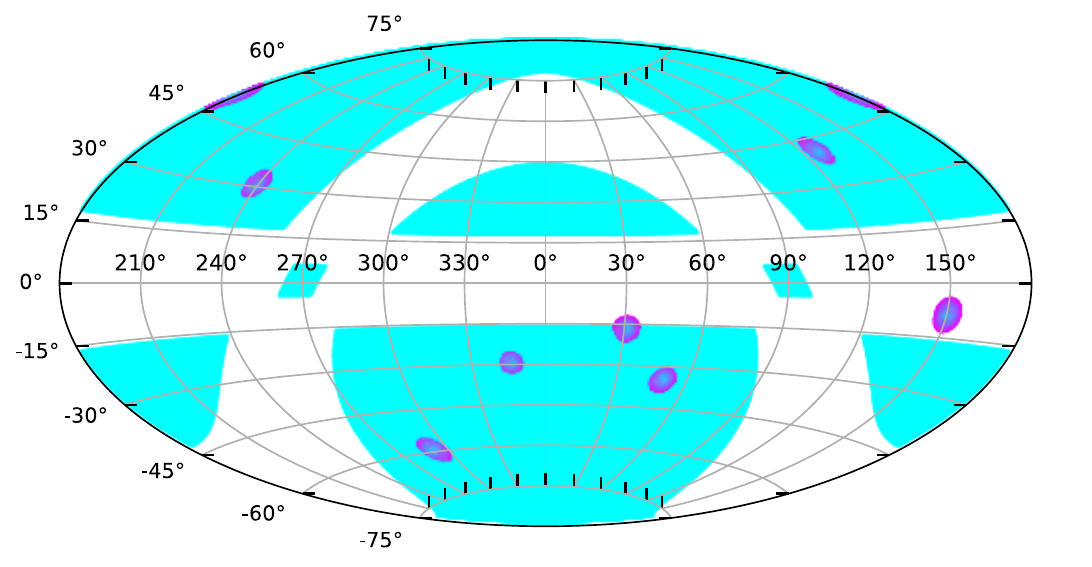}
    \captionsetup{justification=raggedright,width=0.99\linewidth,skip=5pt}
    \caption{The MSC survey strategy is presented in the ecliptic coordinate system. The cyan areas represent the overlapping centers of the CSST wide-field survey observations, while the purple points indicate the centers of the CSST deep-field survey observations. The white regions indicate no observations on the galactic and ecliptic planes.}
    \label{Sec3_msc_pointing}
  \end{minipage}%
  \begin{minipage}[t]{0.5\linewidth}
    \centering
    \includegraphics[scale=0.43]{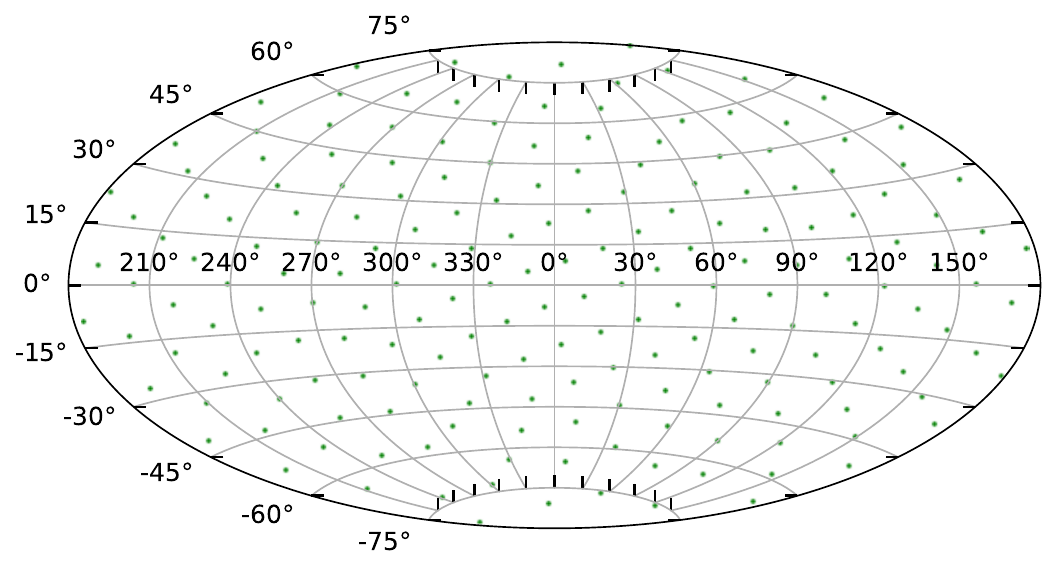}
    \captionsetup{justification=raggedright,width=0.99\linewidth,skip=5pt}
    \caption{The SFP survey strategy is presented in the ecliptic coordinate system. The green points represent the centers of the IFS observations.}
    \label{Sec3_sfp_pointing}
  \end{minipage}
\end{figure}

\subsection{The Mock Survey Strategy Catalog} \label{subsect:planning}
The MSC is a crucial instrument on the CSST, accounting for approximately 70~\% of the total observation time throughout the CSST’s orbital mission. Two types of MSC observations are in detail below:
\begin{itemize}
\item[1.] The CSST's wide-field survey aims to cover over 17,500 square degrees of sky area. Each exposure time must be no less than 150 seconds, and most sky areas will be observed twice for photometry and four times for slitless spectroscopic observation. The CSST survey strategy primarily focuses on these observations. 
\item[2.] The CSST's deep-field survey targets specific sky areas covering no less than 400 square degrees. Each exposure time must be no less than 250 seconds, and most sky areas will be observed four times for photometry and sixteen times for slitless spectroscopic observation. The deep-field survey enables the observation of high-redshift galaxies, quasars, lensed objects, and more.
\end{itemize}

CSST sky survey strategy divides the sky area following the J2000.0 geocentric ecliptic coordinate system, which is conducive to planning the CSST energy balance related to the angle between the solar panels and the Sun. The horizontal direction of the focal plane corresponds to the direction of ecliptic longitude, and the vertical direction of the focal plane corresponds to the direction of ecliptic latitude. 
The size of the sky area is equivalent to the effective FOV of the MSC CCD, which size is 92.32~mm$\times$92.16~mm, corresponding to about $0.1889^{\circ}\times0.1802^{\circ}$ FOV. To facilitate the splicing of sky area images, the overlapping of neighbor sky areas in the vertical direction is at least $10^{\prime\prime}$, and the minimum overlapping area in the horizontal direction is also $10^{\prime \prime}$, which increases with the latitude.

The CSST operates at an altitude of approximately 400 km with an orbital inclination of around $42.5^{\circ}$. It has a precession period of approximately 60 days and an orbital period of about 90 minutes. Various factors in the orbit influence the CSST observation and have been taken into account during the simulation process, including the angle with the solar, moon, and the Earth's bright side, suspend observation when passes the South Atlantic Anomaly (SAA), and reduce the in-orbit maneuvers to avoid the Control Moment Gyroscope (CMG) in high temperatures (\citealt{2019Survey}). 

We plot the mock sky survey strategy in Figure~\ref{Sec3_msc_pointing}. 616,258 MSC center positions are shown, and it is assumed that the CSST scans the sky following these positions in order. The color bar represents the observation times at each point. The area in cyan covers most of the sky, representing the 17500 square degrees wide-field survey. Each MSC center position will be pointed and observed once, and related sky areas should be captured twice due to the two same filters on the MSC. The area in the purple color indicates the deep-field survey. Each MSC center position will be pointed and observed four times at each point, and the related sky areas should be captured in eight images to ensure the depth by overlapping (\citealt{2011Consideration}). 

In addition, 192 observations corresponding to the SFP have been documented in an Additional Mock Survey Strategy Catalog. These observations were generated based on preliminary observation planning to the Integral Field Spectrograph (IFS). Typically, IFS observations involve random rotations around the center. Despite the absence of attitude or quaternion data, the performance of the guide star catalog during SFP observations can still be evaluated to some extent. The spatial distribution of these observation centers is illustrated in Figure~\ref{Sec3_sfp_pointing}.

The mock survey strategy catalog records the details of the mock sky survey strategy, such as the observation ID, central ecliptic coordinate, exposure time, and additional information. This catalog plays an important role in verifying the guide star catalog.

\section{Methods} \label{sect:Method}
A high-quality CSST-MGSC possesses several essential attributes, such as outstanding astrometric accuracy, uniform sky coverage, minimal proper motion, and consistent photometric and astrometric properties. To construct such CSST-MGSC, input catalogs as original material must supply highly precise and uniform astrometric and photometric data for a diverse range of stars across the sky, with negligible systematic errors or biases. As new observations and technological advancements are incorporated, contemporary catalogs are expected to offer enhanced accuracy and comprehensiveness. To further augment the astrometric accuracy of the CSST-MGSC, exceptional sources, like binary stars, should be excluded. Moreover, algorithms designed to maintain uniformity in source density distribution must be employed, considering the specific characteristics of the CSST FGS. The reliable expected instrument magnitude should be derived from the FGS transmission fitting to most guide star types. Finally, following the mock survey strategy catalog, a rigorous validation process should be implemented to identify potential limitations in the CSST observation capabilities.

\subsection{Input Catalogs Preprocessing} \label{subsect:input_catalog}
We selected the $Gaia$ DR3 main source catalog as the input catalog and conducted initial preprocessing. Considering the sensitivity range of FGS spanning magnitudes 8 to 15, a total of 36,846,642 stars were selected based on the $G$-band photometric magnitude criteria. Furthermore, we excluded variable, binary, and high proper motion stars (R.A.PM or Dec.PM $\geq$ 150 mas/yr) by referring to the relevant $Gaia$ DR3 catalogs, aiming to enhance the accuracy of CSST-MGSC. Section \ref{subsect:rs_preprocessing} provides a comprehensive account of the preprocessing methodology and results for the input catalogs.

\subsection{Catalog Index} \label{subsect: catalog index method}
To streamline and accelerate the homogenization process, it is essential to partition the stars in the catalog into equal-area blocks based on their positions and assign them logical indexes. Both the Hierarchical Triangular Mesh (HTM) and the Hierarchical Equal Area isoLatitude Pixelization (HEALPix) are widely adopted and well-established techniques for spatial indexing and partitioning the celestial sphere in astrometry and astrophysics applications. HTM was initially developed by the Sloan Digital Sky Survey (SDSS) team and has since been extensively adopted in various SDSS-related projects. It utilizes a triangular mesh to divide the sphere into a series of triangles, each assigned a unique index. On the other hand, HEALPix was developed by Krzysztof M. Górski and is a hierarchical algorithm that partitions the sphere into equal-area regions with unique indexes arranged in a tree-like structure. 

In comparison, the primary advantage of HEALPix over HTM is its ability to achieve equal-area sky divisions, which greatly facilitates subsequent processing. We have implemented HEALPix to generate a standardized all-sky catalog with the Voronoi algorithm. By setting the HEALPix NSIDE parameter to 32, we partition the celestial sphere into 12,288 quadrangular sky regions, each covering approximately 3.36 square degrees. Such an individual HEALPix sky region is defined as a PIXArea. The sources in an individual HEALPix sky region are saved in separate text files identified by their corresponding HEALPix indices. This results in 12,288 sub-area catalogs, each representing a distinct sky region. All subsequent catalog processing and analysis tasks will be performed using these sub-area catalogs. 

\subsection{Catalog Homogenization}
\label{subsect:homogenization}

%------------------------------------------------------------
\begin{figure}
\centering
\includegraphics[width=0.6\textwidth, angle=0]{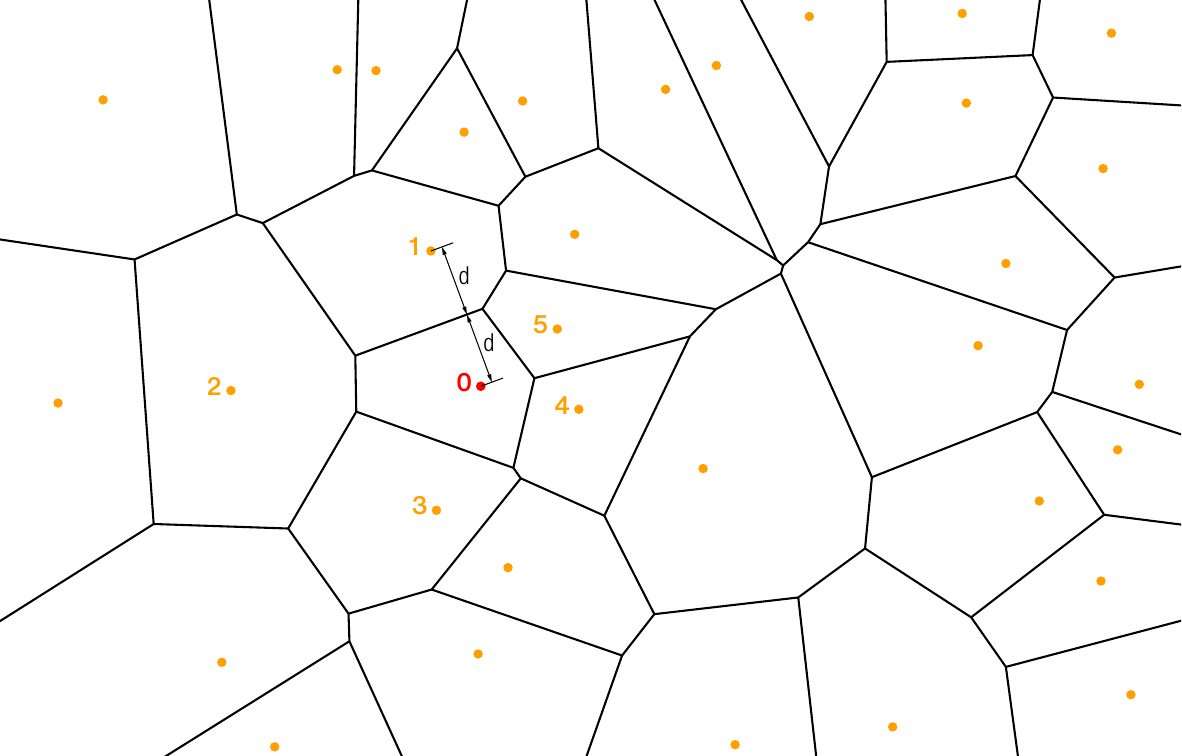}
\caption{The Voronoi algorithm diagram illustrates a partitioning method that divides a given space into distinct regions based on input points, known as seeds or sites. Each region, referred to as a Voronoi cell, contains a single seed (denoted as red "0"), with cell boundaries designed to ensure that points within the cell are closer to their associated seed than any other seeds in the space. This figure depicts the partitioned regions, which consist of vertices, edges, and polygons. The edges, marked with "d" for distance, maintain equidistance between adjacent seeds (indicated as red "0" and orange "2"), while vertices represent points equidistant to three or more seeds (labeled as orange "1", "2", "3", "4", "5").}
\label{Sec3_Fig004}
\end{figure}

%------------------------------------------------------------  
\begin{figure*}
\centering
\includegraphics[width=0.49\textwidth, angle=0]{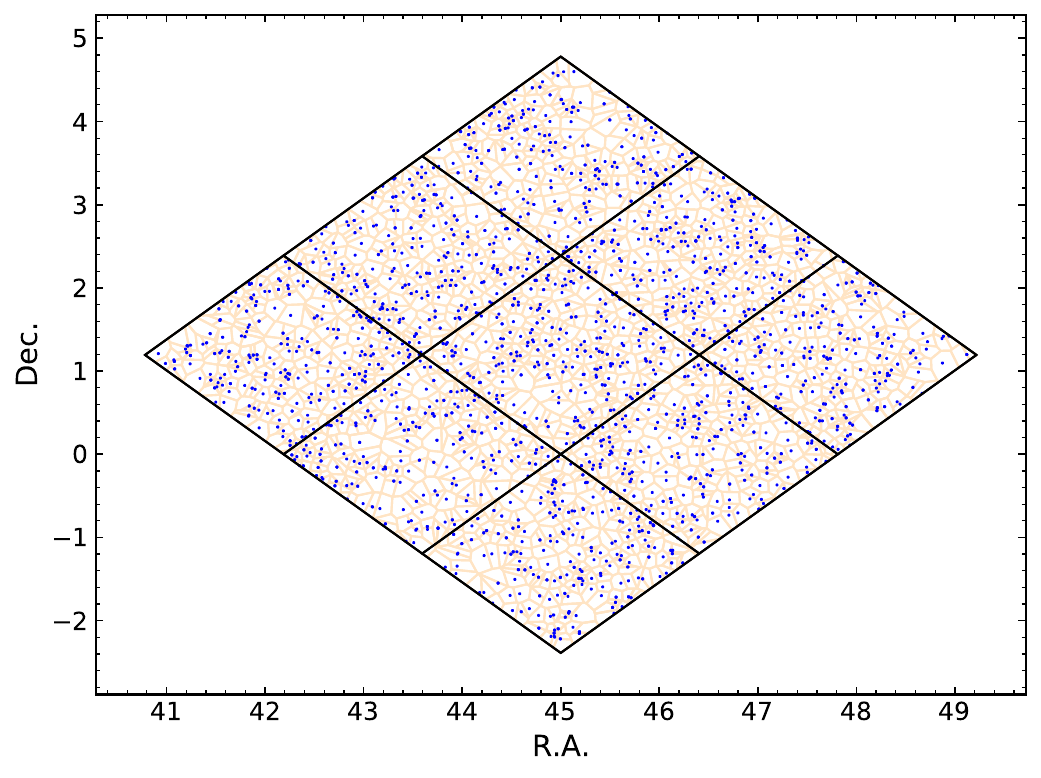}
\includegraphics[width=0.49\textwidth, angle=0]{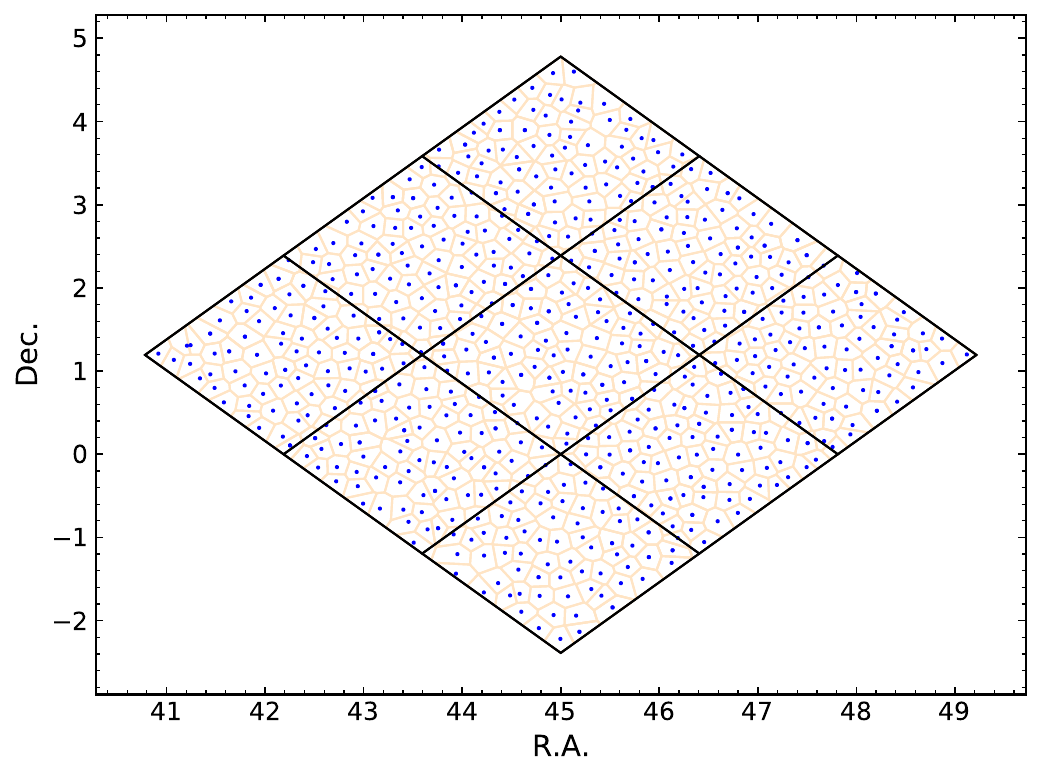}
\caption{Before-and-after homogenization plots showing the distribution of stars.} 
\label{Fig9}
\end{figure*}

A set of uniformly distributed stars in the FOV of the FGS can significantly enhance the accuracy and stability of the positioning and guidance process.
Additionally, it is essential to minimize the catalog size to reduce storage space requirements and data computation complexity. To accomplish this, it is crucial to equalize the star distribution in the guide star catalog across diverse sky regions by iteratively eliminating stars from crowded fields until a consistent number persists within each region.  

The Voronoi algorithm, also known as Voronoi diagram or Voronoi tessellation, is a computational geometry technique that partitions a space into regions based on the proximity to a set of input points. As Figure~\ref{Sec3_Fig004} shows, each region corresponds to the area closest to a specific input point, and the boundary between two regions corresponds to the points equidistant to the two nearest input points. The Voronoi diagram has numerous applications in pattern recognition, computer graphics, and geographic information systems. Refer to \cite{2016A}, we create a homogenization method based on the Voronoi algorithm in the density stars PIXArea. First, we build the Voronoi diagram in the input catalog's PIXArea. Then, we remove stars from the smallest area of the Voronoi cell and rebuild the Voronoi diagram again. This procedure will run iteratively until there is a fixed number of stars in each PIXArea.

Figure \ref{Fig9} presents an example of star distribution before and after homogenization. The left panel illustrates the star distribution within the 9 PIXAreas before homogenization. In contrast, the right panel displays the outcome after homogenization, with 84 stars within the central PIXArea. Further descriptions of the homogenization results are presented in Section \ref{subsect:rs_homog}.

\subsection{The Expected Instrument Magnitude} \label{subsect:instmag}

%------------------------------------------------------------
\begin{figure}
\centering
\includegraphics[width=0.9\textwidth, angle=0]{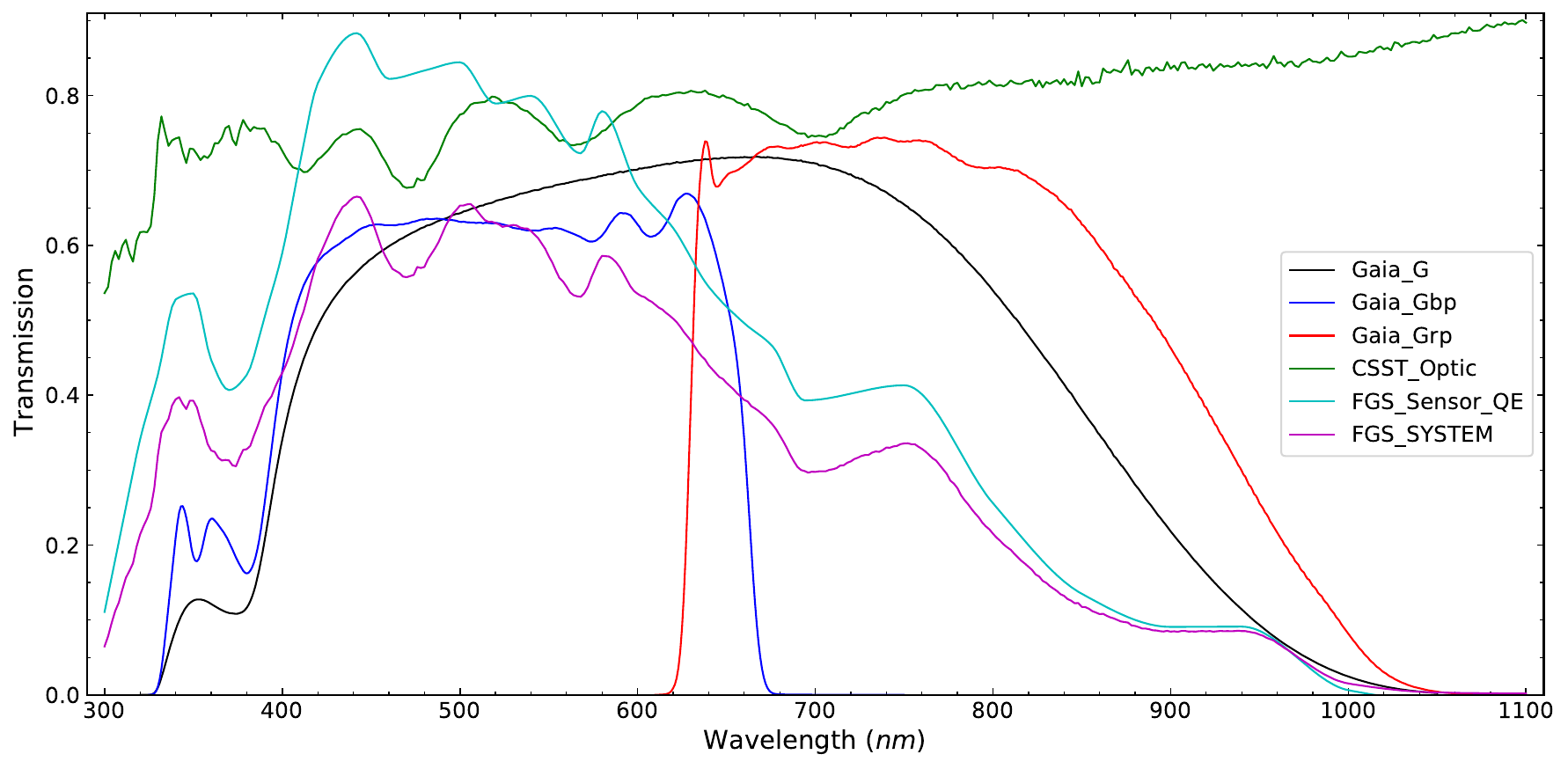}
\caption{The $Gaia$ and CSST transmissions. In the figure, the black line represents the transmission of $Gaia$ G, the blue line represents the transmission of $Gaia$ BP, and the red line represents the transmission of $Gaia$ RP. The green line represents CSST expected optical transmission, the cyan line represents FGS expected sensor's QE, and the purple line represents FGS system transmission, combined with sensor and optical transmissions\protect\footnotemark.}
\label{Sec3_Fig001}
\end{figure}
\footnotetext{http://svo2.cab.intacsic.es/theory/fps/index.phpmode=browse\&gname=GAIA\&gname2=GAIA3\&asttype=}

%------------------------------------------------------------
\begin{figure}
\centering
\includegraphics[width=0.9\textwidth, angle=0]{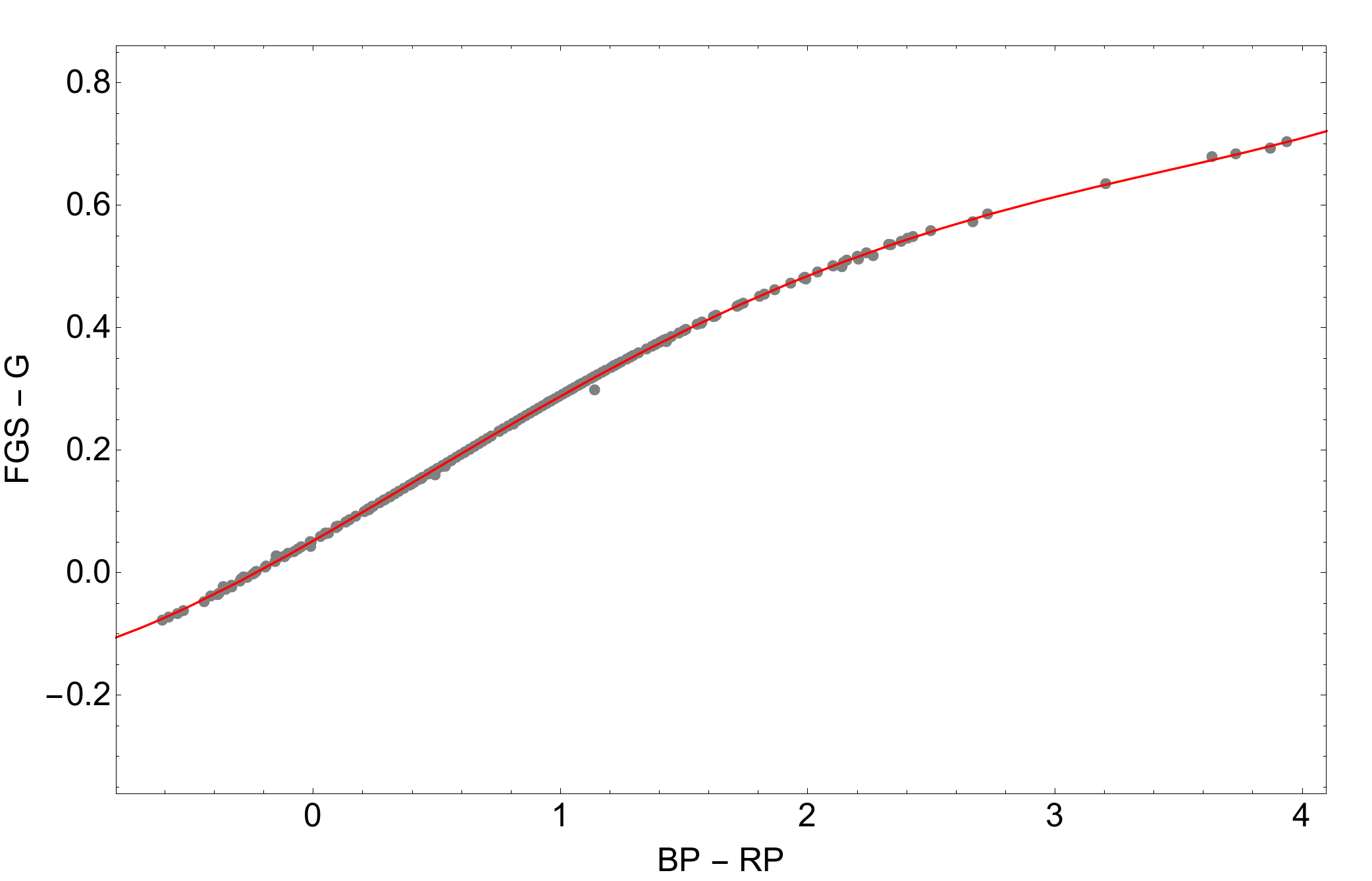}
%\captionsetup{justification=raggedright,width=\linewidth,skip=5pt}
\caption{The fourth-order polynomial function, denoted as Eq.~\eqref{eq6}, translates the $Gaia$ G magnitude (${G}$) into the FGS instrument magnitude ($MAG_{FGS}$). In the figure, the X-axis represents the color index (${BP}$ - ${RP}$), while the Y-axis represents the color index ($MAG_{FGS}$-${G}$). Each blue point on the plot represents the relationship between the two color indexes, which is derived from convolutions of one $LAMOST$ template spectrum and the system transmissions of FGS, G, BP, and RP, respectively. The red line on the plot represents the fourth-order polynomial fitting of these points. The residual of the fitting is 0.0026.}
\label{Sec3_instmag}
\end{figure}

The transmissions for $Gaia$ and CSST are illustrated in Figure~\ref{Sec3_Fig001}, depicting the same wavelength range. The FGS sensor has an expected quantum efficiency (QE) of up to 89\%. Its most effective range is within the visual band, spanning from 400 nm to 760 nm. However, considering the expected optical transmission of CSST, the maximum transmission of the FGS system decreases to 66.5\%. The system transmission of the FGS is comparable to the $Gaia$ $G$ band within the range of 300 nm to 550 nm, but the difference becomes more pronounced as it approaches 700 nm. Then, the difference gradually diminishes as the wavelength approaches 1000 nm.  

The color index is linked to a star's effective temperature and the photometric system's transmission, which could be described in Eq.~\eqref{eq5}. In Eq.~\eqref{eq5}, the $MAG_{1}$ and $MAG_{2}$ represents one star observed by different instruments in bands [$\lambda_2$,$\lambda_1$] and [$\lambda_4$, $\lambda_3$]. The observed star's effective temperature is $T$, and the star's flux related to the wavelength is represented by $Flux(\lambda, T)$. Two observation instruments' transmissions are represented by $Trans_1(\lambda)$ and $Trans_2(\lambda)$.
In theory, the color index, which corresponds to the FGS instrument magnitude ($MAG_{FGS}$) and $Gaia$ G, can be linearly represented by the (BP-RP) color index. Combined with the transmission data presented in Figure~\ref{Sec3_Fig001}, we can establish this relationship by fitting different types of stars with specific effective temperatures. 
\begin{equation}\label{eq5}
\begin{split}
   Color\_Index &= MAG_{1} - MAG_{2}\\
   &= -2.5\lg\frac{\int_{\lambda_1}^{\lambda_2}Flux(\lambda,T)\cdot Trans_1(\lambda)d(\lambda)}{\int_{\lambda_3}^{\lambda_4}Flux(\lambda,T)\cdot Trans_2(\lambda)d(\lambda)}
\end{split}
\end{equation}

To achieve this, we employ the $LAMOST$ Stellar Classification Template, which consists of 61 spectra types, encompassing O-type and B-type spectra, A-type spectra, F-, G-, and K-type spectra, M-type spectra, as well as other types of spectra, such as Carbon Star, White Dwarf (WD), Carbon WD, Magnetic WD, Double Star, and Cataclysmic Variable (\citealt{2014AJ.147.101W}). 
We can establish a relationship between the two color indices by convoluting one template spectrum and the system transmissions of FGS, G, BP, and RP. 
As a result, 61 spectra types could create 61 points in the color-color map. The fourth-order polynomial fitting is adapted to these data, shown in Figure~\ref{Sec3_instmag}. In the figure, the X-axis represents the color index (${BP}$ - ${RP}$), and the Y-axis represents the color index ($MAG_{FGS}$ - $G$). 
The fitted fourth-order polynomial function $g(x)$, detailed in Eq.~\eqref{eq6}. And the residual of the fitting is 0.0026.    
\begin{equation}\label{eq6}
\begin{split}
   &x = {BP} - {RP}\\
   &g(x) = 0.0513035 + 0.230806\cdot x + 0.0234848\cdot x^2 - 0.0205912\cdot x^3 + 0.00264465\cdot x^4
\end{split}
\end{equation}

There are several issues with this approach. Firstly, there are unknown discrepancies between the expected and actual transmission, and the actual transmission needs to be derived from laboratory testing and calibration observations, which have not yet been finalized. Secondly, the current method based on template spectra has a shorter wavelength range than the unfiltered sensor or the $G$-band, resulting in potential calculation errors. Thirdly, the available template spectra are limited and do not encompass extra-galactic sources, leading to deviations between the fitting results and the actual translation. However, the expected FGS instrument magnitude can still play a significant role in the sky survey strategy study, FGS image simulation, and the verification of the star processing algorithm. This parameter will be published in the CSST-MGSC as the "phot\_inst\_mean\_mag" column, as described in Table~\ref{Tab001}.

\subsection{Catalog Validation} \label{subsect:GnoPro}
For each observation center in the mock survey strategy catalog, we should identify the guiding stars in the FGS sensor's FOV, and accurately convert their celestial coordinates to the sensor's measurement coordinates. Based on the classic theory of astrometry, a star on the celestial sphere is imaged on a point on the theoretical focal plane of the telescope. Still, the radiation of the star is not directly received by the theoretical focal plane, but received by one or more physical media surfaces, such as the receiving surface of a CCD or CMOS deployed at the focal plane. In an ideal situation, these physical surfaces together constitute a tangent plane of the theoretical focal plane, and the position of the star image needs to be described on this tangent plane. To this end, one or more measurement coordinates corresponding to physical medium surfaces must be constructed on this plane. The direction and scale of the measurement coordinate system are determined by the physical parameters of the physical media (sensor's pixel, for example). To facilitate the star image position measurement, it is also necessary to construct a theoretical coordinate, which has a similar form to the measurement coordinate and a definite transformation relationship with the celestial coordinate defined by the projection mode. The theoretical coordinate is often called the standard coordinate.

The objective of validating the guide star catalog involves using the mock survey strategy catalog and the CSST-MGSC to ascertain the feasibility of specific mock observations in the Mock Survey Strategy Catalog, mainly when guide stars are adequate. Consequently, it is essential to compute the number of stars the FGS FOV encompasses. 
Taking the MSC for example, we selected an observation center $(A_O, D_O)$ from the Mock Survey Strategy Catalog to construct the standard coordinate system, with the coordinate center assumed to be at $T(A_O, D_O)$. Employing a spherical arc length of 1.8$^{\circ}$, as derived from the formula presented in Eq.~\eqref{eq3}, we proceeded to identify all sources within the MSC FOV using the CSST-MGSC. The standard coordinates of these sources were then determined based on the gnomonic projection equations detailed in Eq.~\eqref{eq1} and Eq.~\eqref{eq2}. Subsequently, concerning the corners of each CMOS in the standard coordinate, as delineated in Table~\ref{Sec3_Tab002}, we searched for guide stars within the FOV of each CMOS, accounting for edge effects. The number of available stars must surpass the FGS requirement; otherwise, the observation accuracy will be compromised, or the observation may fail altogether.
\begin{equation}\label{eq3}
   S = R~\arccos[\cos\beta~\cdot \cos{D_O}\cdot~\cos(\lambda - {A_O}) + \sin\beta~\cdot \sin{D_O}]
\end{equation}
where $R$ is the celestial sphere radius that we could make as unit one; $(\lambda,\beta)$ is the ecliptic coordinate of a star in CSST-MGSC, and $(A_o, D_o)$ is the ecliptic coordinate of the observation center from the Mock Survey Strategy Catalog. 

The validation of the guide star catalog on the SFP is similar to that on the MSC. The coordinates of each CMOS corner in the standard coordinate system are described in Table~\ref{Sec3_Tab003}.

\section{Results}   \label{sect:Res} 
In this section, we present the analysis and processing results following the methods mentioned above. Initially, the $Gaia$ DR3 main source catalog was used as the input catalog with certain limitations, such as magnitude range and exclusion of high proper motion stars, binaries, and variables. We carefully considered the impact of different types of objects on the catalog and selectively retained specific types of objects for certain observations to enhance the performance of the FGS. Next, we conducted an analysis and applied a homogenization procedure to reduce the size of the catalog while ensuring that the number of stars captured by the FGS was not significantly reduced. We also optimized the FGS operation mode in different sky regions based on the mock results. Finally, we tested the original version of CSST-MGSC using the all-sky mock survey strategy and evaluated its performance to verify whether it met the FGS requirements set by the CSST mission engineering team.

\subsection{Input Catalogs Preprocessing Results} 
\label{subsect:rs_preprocessing}

%------------------------------------------------------------
\begin{figure}
  \begin{minipage}[t]{0.5\linewidth}
    \centering
    \includegraphics[scale=0.5]{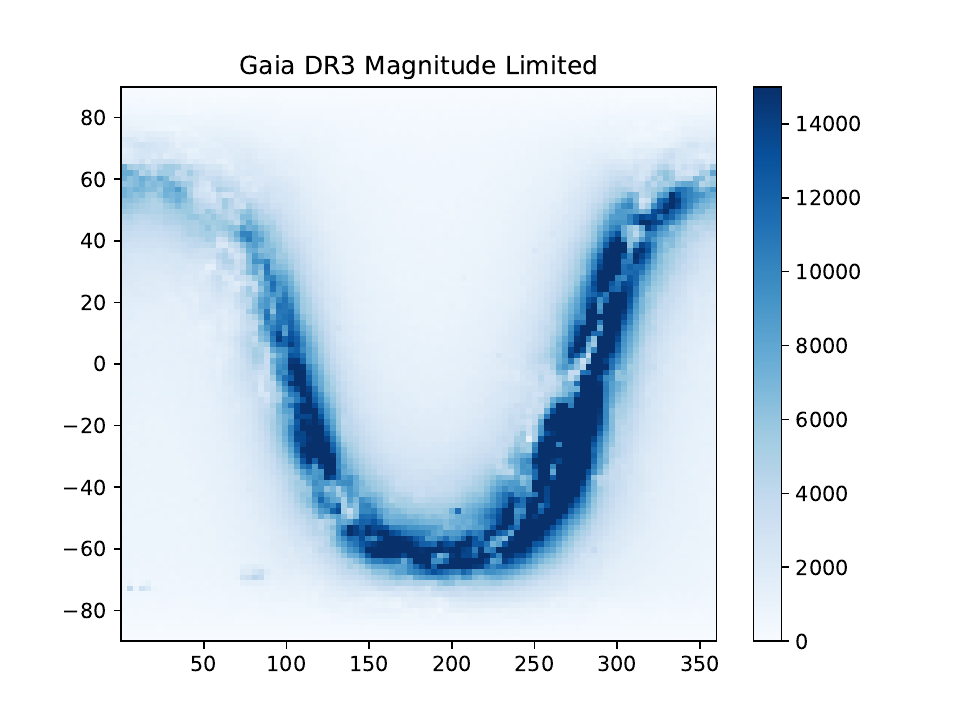}
    \captionsetup{justification=raggedright,width=0.99\linewidth,skip=5pt}
    \caption{The density distribution of the $Gaia$ DR3 main source catalog in the equatorial coordinate system, has limited magnitudes ranging from G=8 to 15 and proper motions less than 150 mas/yr. The axis unit is degree. }
    \label{Sec5_gaiadr3}%\label{Fig006_1}
  \end{minipage}
  \begin{minipage}[t]{0.5\linewidth}
    \centering
    \includegraphics[scale=0.5]{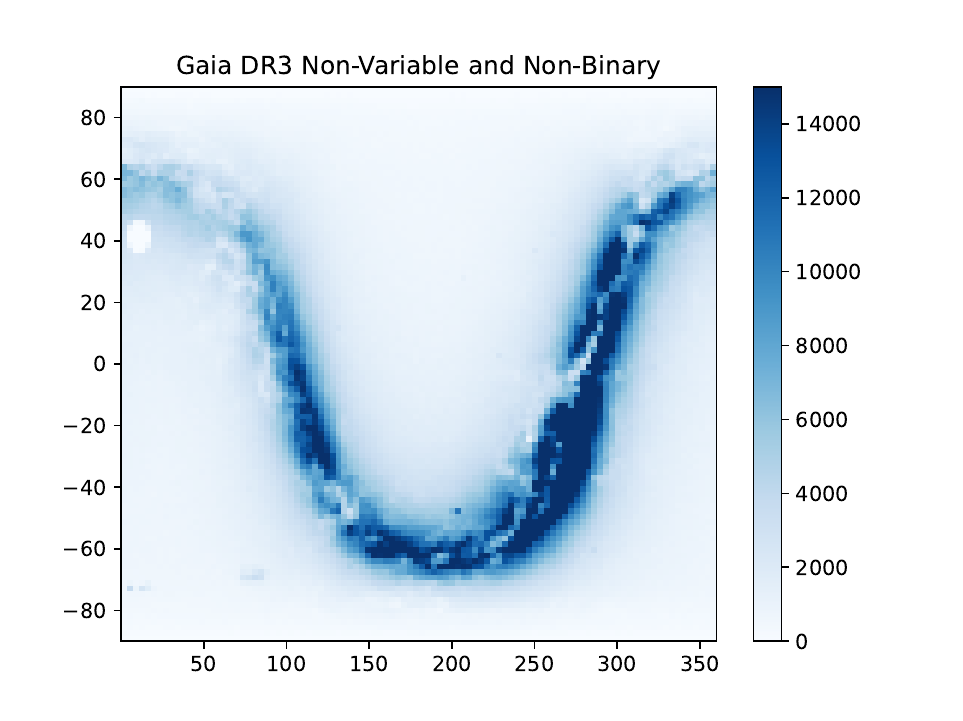}
    \captionsetup{justification=raggedright,width=0.99\linewidth,skip=5pt}
    \caption{The density distribution of the $Gaia$ DR3 main source catalog in the equatorial coordinate system, has limited magnitudes ranging from G=8 to 15 and proper motions in R.A. and Dec. less than 150 mas/yr, after excluding binary and variable stars. The axis unit is degree.}
    \label{Sec5_gaiadr3_nobnov}%\label{Fig006_4}
  \end{minipage}
\end{figure}

%------------------------------------------------------------fig 17/18
\begin{figure}
  \begin{minipage}[t]{0.5\linewidth}
    \centering
    \includegraphics[scale=0.49]{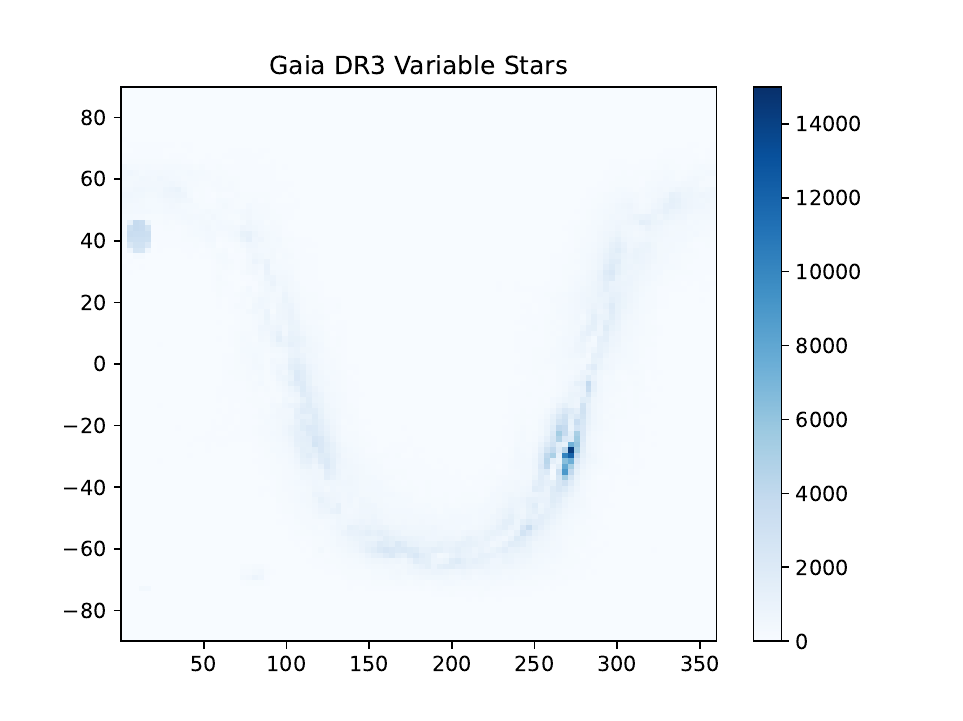}
    \captionsetup{justification=raggedright,width=0.99\linewidth,skip=5pt}
    \caption{The density distribution of the $Gaia$ DR3 variable catalog in the equatorial coordinate system, has limited magnitudes ranging from G=8 to 15 and proper motions in R.A. and Dec. less than 150 mas/yr (\citealt{Eyer2022}). The axis unit is degree.}
    \label{Sec5_gaiadr3_var} %\label{Fig006_2}
  \end{minipage}%
  \begin{minipage}[t]{0.5\linewidth}
    \centering
    \includegraphics[scale=0.5]{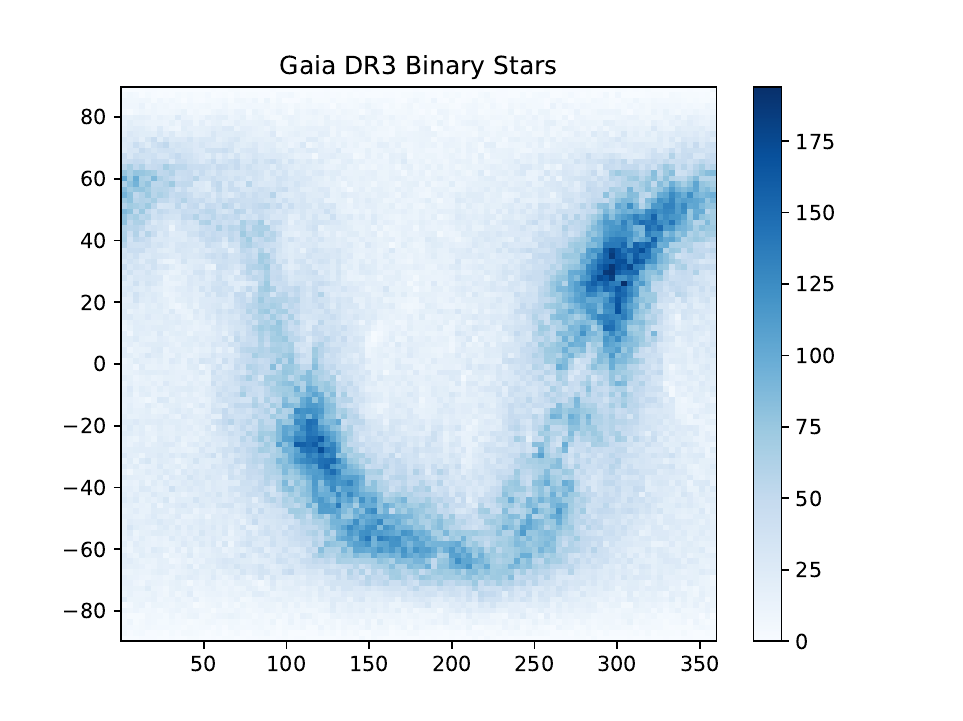}
    \captionsetup{justification=raggedright,width=0.99\linewidth,skip=5pt}
    \caption{The density distribution of binary in the equatorial coordinate system,  combined with four binary catalogs, has limited magnitudes ranging from G=8 to 15 and proper motions in R.A. and Dec. less than 150 mas/yr (\citealt[Siopis et al, 2023, in prep; Damerdji et al., 2023, in prep] {Halbwachs2022, Holl2022}). The axis unit is degree.}
    \label{Sec5_gaiadr3_binary}%\label{Fig006_3}
  \end{minipage}
\end{figure}

%------------------------------------------------------------
\begin{figure}
\centering
\includegraphics[width=0.48\textwidth, angle=0]{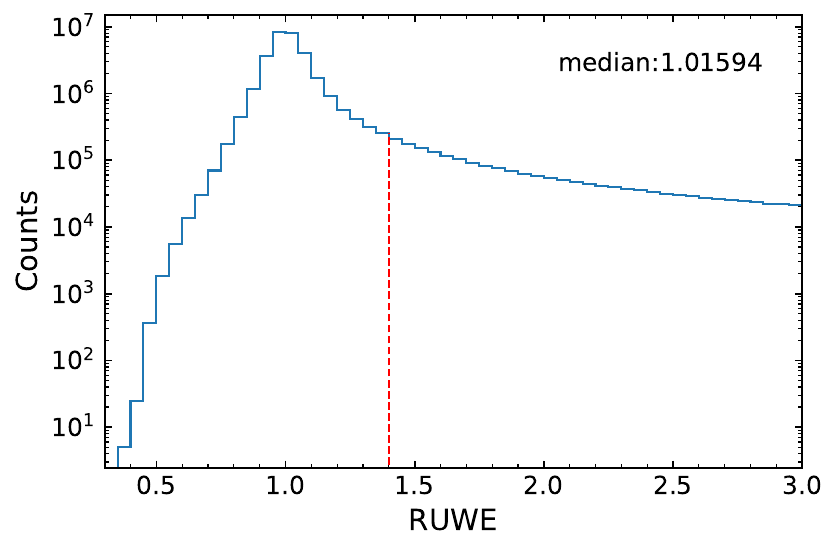}
\includegraphics[width=0.48\textwidth, angle=0]{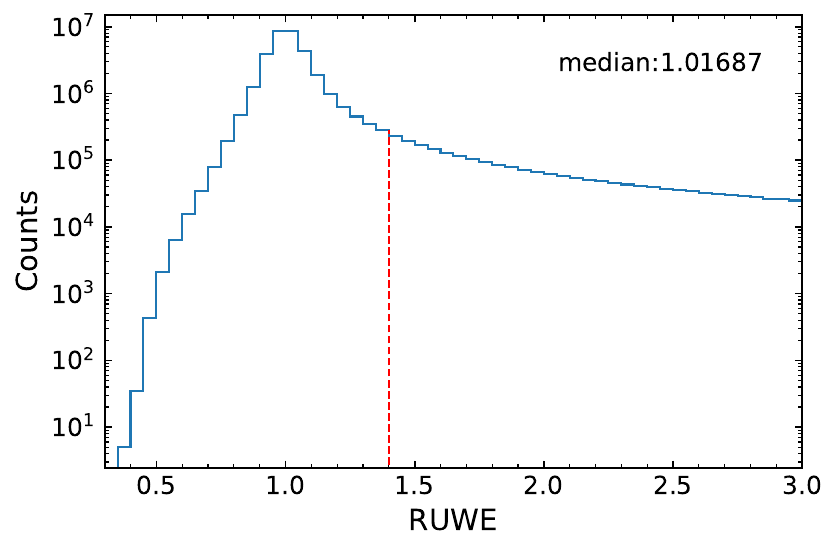}
% \captionsetup{justification=raggedright,skip=5pt}
\caption{The RUWE of $Gaia$ DR3 without binaries and variables ($left$) or without binaries ($right$). The RUWE distributions without and with variables are almost the same, and the error imported by variables is slight. The values of RUWE of the left and the right all range from 0.37725 to 116.01637, and their Q90 are 1.42199 and 1.46424, respectively. The red dotted lines mark the value of 1.4, which is regarded as the typical data quality threshold value for Gaia (\citealt{pearce2022ruwe}). }
\label{Sec5_gaiadr3_ruwe} %\label{Sec3_Fig002}
\end{figure}

%________________________________________ 
\begin{table}\centering
\begin{minipage}[]{\textwidth}
\caption{Comparisons of errors and RUWEs for binary, variable, high proper motion, and other point-like objects in $Gaia$ DR3 with $G$ magnitude range 8-15 mag.
\label{Sec3_Tab001}} \end{minipage} \\
\setlength{\tabcolsep}{1.5mm}
\small
 \begin{tabular}{ccccccc}
  \hline
  \hline\noalign{\smallskip}
Catalog & Name & Q10 & Q50 & Q75 & Q90 & Quantity \\
  \hline
  \hline\noalign{\smallskip}
\multirow{6}{*}{Binary} & ra\_err & 0.01 & 0.03 & 0.04 & 0.06  & 473 794 \\
~ & dec\_err & 0.01 & 0.02 & 0.04 & 0.06  & 473 794 \\
~ & pmra\_err & 0.02 & 0.03 & 0.05 & 0.08  & 473 794 \\
~ & pmdec\_err & 0.02 & 0.03 & 0.05 & 0.07  & 473 794 \\
~ & ruwe & 1.45 & 2.16 & 3.11 & 4.72  & 473 794 \\
~ & phot\_g\_mean\_flux\_err & 6.28 & 22.21  & 60.04 & 185.97 & 473 794 \\
\hline
  \multirow{6}{*}{Variable} & ra\_err & 0.01 & 0.02 & 0.03 & 0.06  & 2 384 556 \\
~ & dec\_err & 0.01 & 0.02 & 0.03 & 0.05  & 2 384 556 \\
~ & pmra\_err & 0.01 & 0.02 & 0.04 & 0.07  & 2 380 729 \\
~ & pmdec\_err & 0.01 & 0.02 & 0.03 & 0.06  & 2 380 729 \\
~ & ruwe & 0.92 & 1.03 & 1.14 & 1.8  & 2 380 729 \\
~ & phot\_g\_mean\_flux\_err & 8.76 & 35.48  & 137.07 & 547.35 & 2 384 556 \\
\hline
\multirow{6}{*}{High PM} & ra\_err & 0.01 & 0.02 & 0.02 & 0.05  & 14 266 \\
~ & dec\_err & 0.01 & 0.01 & 0.02 & 0.05 & 14 266 \\
~ & pmra\_error & 0.01 & 0.02 & 0.03 & 0.07 & 14 266 \\
~ & pmdec\_error & 0.01 & 0.02 & 0.03 & 0.06 & 14 266 \\
~ & ruwe & 0.96 & 1.13 & 1.30 & 2.72 & 14 266 \\
~ & phot\_g\_mean\_flux\_err & 9.94 & 42.49 & 177.89 & 570.83 & 14 266 \\
\hline
\multirow{6}{*}{Other Stellar} & ra\_err & 0.01 & 0.02 & 0.02 & 0.03  & 33 974 026 \\
~ & dec\_err & 0.01 & 0.01 & 0.02 & 0.03  & 33 974 026 \\
~ & pmra\_err & 0.01 & 0.02 & 0.03 & 0.04  & 33 704 553 \\
~ & pmdec\_err & 0.01 & 0.02 & 0.02 & 0.03  & 33 704 553 \\
~ & ruwe & 0.93 & 1.02 & 1.09 & 1.42  & 33 704 553 \\
~ & phot\_g\_mean\_flux\_err & 5.04 & 10.09  & 23.99 & 68.75 & 33 974 026 \\
  \noalign{\smallskip}\hline
  \hline
\end{tabular}
\tablecomments{\textwidth}{The "High PM" refers to stars with R.A. or Dec proper motion $\geq 150$ mas/yr. The "Other Stellar" refers to sources that are not binary, variable, and high proper motion stars. }
\end{table}

%------------------------------------------------------------
\begin{table}[t]
\setlength{\tabcolsep}{1mm}
    \caption{Correlational statistics of two input source catalogs within the 8-15 magnitude at $G$-band.} 
    \label{table8}
    \begin{minipage}{0.5\textwidth}
      \centering
        (a) Preprocessed $Gaia$ DR3 Catalog
        \resizebox{7.25cm}{1.75cm}{
        \small
        \begin{tabular}{cccc}
            \hline
  \hline\noalign{\smallskip}
 Magnitude &  Ratio & Population & Percent   \\
 (mag) & ($N_{\rm{star}}$/deg$^{2}$) & &($N_{\rm{star}}$/$N_{\rm{total}}$) \\
  \hline
  \hline\noalign{\smallskip}
8.0 $\le G < $ 10.0 & 20.41 & 357 212 & 1.05 \\
10.0 $\le G < $ 12.0 & 133.57 & 2 337 530 & 6.88 \\
12.0 $\le G < $ 14.0 & 712.04 & 12 460 704 & 36.68 \\
14.0 $\le G < $ 15.0 & 1075.35 & 18 818 580 & 55.39 \\
  \noalign{\smallskip}\hline
  \hline
        \end{tabular}
        }
    \end{minipage}%
    \begin{minipage}{0.5\textwidth}
      \centering
        (b) Preprocessed $Gaia$ DR3 Catalog Added Variables
        \resizebox{7.25cm}{1.75cm}{
        \begin{tabular}{cccc}
            \hline
  \hline\noalign{\smallskip}
 Magnitude &  Ratio & Population & Percent   \\
 (mag) & ($N_{\rm{star}}$/deg$^{2}$) & &($N_{\rm{star}}$/$N_{\rm{total}}$) \\
  \hline
  \hline\noalign{\smallskip}
8.0 $\le M < $ 10.0 & 22.74 & 397 866 & 1.09 \\
10.0 $\le M < $ 12.0 & 144.17 & 2 522 953 & 6.92 \\
12.0 $\le M < $ 14.0 & 773.78 & 13 541 134 & 37.16 \\
14.0 $\le M < $ 15.0 & 1141.36 & 19 973 863 & 54.82 \\
  \noalign{\smallskip}\hline
  \hline
        \end{tabular}
        }
    \end{minipage} 
\tablecomments{\textwidth}{The column 'Ratio’ provides the ratio of the stars' population to the areas (17500 deg$^{2}$) of the focal plane in different magnitude ranges; 'Population' represents stars' population in different magnitude intervals; 'Percent' shows stars' proportion in the total number of stars in corresponding magnitude intervals.}
\end{table}

The $Gaia$ main source catalog is known for its high accuracy in both photometric and astrometric measurements, making it an excellent choice as the input catalog for building the CSST-MGSC. However, to further enhance the astrometric accuracy of the CSST-MGSC, special attention is paid to three types of sources: binary stars, variable stars, and stars with high proper motion. These particular sources require in-depth analysis to understand how they can potentially impact the accuracy of the star catalog, using the $Gaia$ DR3-related catalogs.

In the case of binary stars, the orbital motion of the constituent stars around their common center of mass results in variations in their relative positions as observed from Earth. The elliptical nature of binary star orbits leads to periodic position fluctuations correlated with the orbital parameters. Additionally, for closer binary systems, the light from a binary star system may appear as a single point with an irregular shape in the telescope image, making it difficult to accurately determine the individual stars' positions. Consequently, the astrometric positions of binary stars tend to have higher uncertainties than those of single stars. Therefore, the binary stars should be removed from the output catalog.

For variable stars, their brightness changes over time, which can result in differences between the observed flux measured by the FGS and the cataloged values. This can potentially impact the accuracy of CSST guiding, as the chosen exposure time may be incorrect due to the variability in flux, and the signal-to-noise ratio (S/N) of centroiding may be low.
Two origins of variables, classification and specific object studies (SOS), are present in the $Gaia$ archive. We chose the classification method due to its higher completeness and inclusion of a broader range of variable types. $Gaia$ classification identifies a total of 24 types of variables. However, we focus on variables with the peak-to-peak semi-amplitude in magnitude (TR) median values (Q50) above 0.1 mag, which corresponds to approximately 10\% flux change. Consequently, we narrowed the selection to only 11 types of variables, including Active Galactic Nuclei (or QSOs), Cepheids, Cataclysmic variables, Eclipsing binaries (EBs), Long-period variables (LPVs), Microlensing events, R Coronae Borealis stars, RR Lyrae stars, Short time scale, Supernovae, Symbiotic System. 
Long-period variables (LPVs), Eclipsing binaries (EBs), and Active Galactic Nuclei (or QSOs) are the three majority Variable types with different variable causes. Their proportions are about 36\%, 34\%, and 16\%.
LPVs have been known and studied for a long time due to their large variability amplitudes in the visual band and observations in an extensive volume of space. Their observable features represent the late evolutionary stages of low- and intermediate-mass stars (\citealt{2023A&A...674A..15L}). 
QSOs could be observed by strong flux variations at all wavelengths and an optical continuum containing no emission features (\citealt{1993ApJS...87..451H}). As point-like extragalactic objects, QSOs could used to align Gaia-CRF3 to ICRF \citep{Lebzelter2022} with high astrometric accuracy. 
EBs are small fractions among all the binary stars, revealed as eclipsing, closely aligned to the line of sight (\citealt{2006AAS...209.0604G}), which could be used as a standard candle or the black hole candidates (\citealt{2020ApJS..249...31Y}). Compared to the LPVs, the EBs manifest an extrinsic variability.

To evaluate the impact of variable sources on catalog precision, we compared the statistical results of these parameters among binaries, variables, high proper motion stars, and the remaining stars in our catalog. The summarized results are presented in Table \ref{Sec3_Tab001}. 
The variables' position and proper motion median errors were generally lower than those for binaries but higher than the objects in the "Other Stellar" category. The RUWE values for variables are generally lower than those for binaries and high proper motion stars, but a little higher than for stars in the "Other Stellar" category. The photometric flux mean error at the $G$ band is higher than binaries and stars in the "Other Stellar" category. This should be increased due to the variable nature of multiple observations. In addition, the variables in G, BP, and RP magnitudes affect the derived magnitude following the method introduced in \ref{subsect:instmag}. Adopting the variables will increase the instrument magnitude computation complexity and errors. 

$Gaia$ DR3 variables catalog contains the $Gaia$ Andromeda Photometric Survey (GAPS). GAPS encompasses all 1 257 319 sources within a cone of 5.5 degrees opening directed towards the Andromeda galaxy. 
A different approach to detect variables is employed within M31, which involves examining the correlations between the three passbands in the residuals relative to the mean for each source using principal component analysis (PCA). 12,618 sources have been identified and flagged as variables using this method, accounting for approximately 1\% of all GAPS stars. 
These variables include $\delta$ Scuti stars, $\gamma$ Doradus stars, RS Canum Venaticorum-type variables, long-period variable stars (Miras, semi-regular, OGLE small amplitude red giants), ZZ Ceti variables, as well as certain types of variability associated with binaries\citep{evans2023gaia}. 

High proper motion stars pose a hidden danger in engineering. Improper handling of the proper motion can result in significant position deviations, leading to guide star mismatch. Furthermore, due to the closer distance of the stars with high proper motion, they tend to be brighter and are more likely to be selected for guiding, increasing the likelihood of encountering this issue. 
Furthermore, an analysis of Table \ref{Sec3_Tab001} reveals that high proper motion stars exhibit similar mean positions, proper motion errors, and RUWE values compared to the "Other Stellar" category in Q10, Q50, and Q75. However, these values become higher in Q90. Notably, the photometric mean error is the largest among all the categories.

Considering the minimum quantity among all four types of stars, we have excluded high proper-motion stars from our catalog. This exclusion will not significantly affect the stellar density of the catalog.

Based on these analyses, we initiated the processing of the input source catalogs. Firstly, we applied restrictions to the $Gaia$ DR3 main source catalog, limiting the magnitude range to 8-15 mag and R.A. or Dec. proper motion to less than 150 mas/yr. The resulting density distribution of the processed catalog is illustrated in Figure \ref{Sec5_gaiadr3}. Secondly, we removed variable stars and binary stars from this catalog. The density distribution of the processed catalog without variables and binaries is shown in Figure \ref{Sec5_gaiadr3_nobnov}. The density distributions of variable stars and binary stars, with the same magnitude and proper motion limitations, are depicted in Figure \ref{Sec5_gaiadr3_var} and Figure \ref{Sec5_gaiadr3_binary}, respectively. 

We have summarized the average number of stars per square degree in each magnitude bin following the CSST mock survey strategy to compare the guide star catalog performances with or without variable stars. 
Tab.~\ref{table8}(a) represents the results without including variables, while Tab.~\ref{table8}(b) includes variables. From these tables, it can be inferred that after including the variable stars, there is a slight increase in the number of stars in each magnitude bin. Meanwhile, the distribution of magnitudes shifts towards the brighter end by approximately 0.5\% in the last three magnitude bins. Furthermore, we examined the RUWE for the same sky area, both with and without including variables. Figure \ref{Sec5_gaiadr3_ruwe} shows that the RUWE slightly increases when variables are included, indicating a small error introduced by the variable stars. Based on these findings, it is recommended to exclude variable stars to ensure higher accuracy. As Figure \ref{Sec5_gaiadr3_nobnov} shows, due to the inclusion of GAPS data in the $Gaia$ Variable Catalog, removing variable stars results in no stars being left in the M31 region. In light of the significance of M31 as an observation area, we have initiated a recall of all objects within this region based on their magnitude range and variable types. The selection and optimization of guide stars in this area are still under investigation.

Consequently, we removed 2,384,556 objects from the variable catalog (\citealt{Eyer2022}) and 473,794 objects from the four binary catalogs (\citealt[Siopis et al, 2022, in prep; Damerdji et al., 2022, in prep]{Halbwachs2022, Holl2022}). It remains over 33,988,292 stars in the catalog. Furthermore, we excluded 14,266 high proper motion stars with R.A. or Dec. proper motion value greater than or equal to 150 mas/yr. Lastly, the processed $Gaia$ DR3 main source catalog contains 33,974,026 stars, with the magnitude distribution depicted in Figure~\ref{Fig006_5}($left$).

\subsection{Results of Catalog Homogenization} 
\label{subsect:rs_homog}

%------------------------------------------------------------
\begin{figure}
\centering
\includegraphics[width=1.0\textwidth, angle=0]{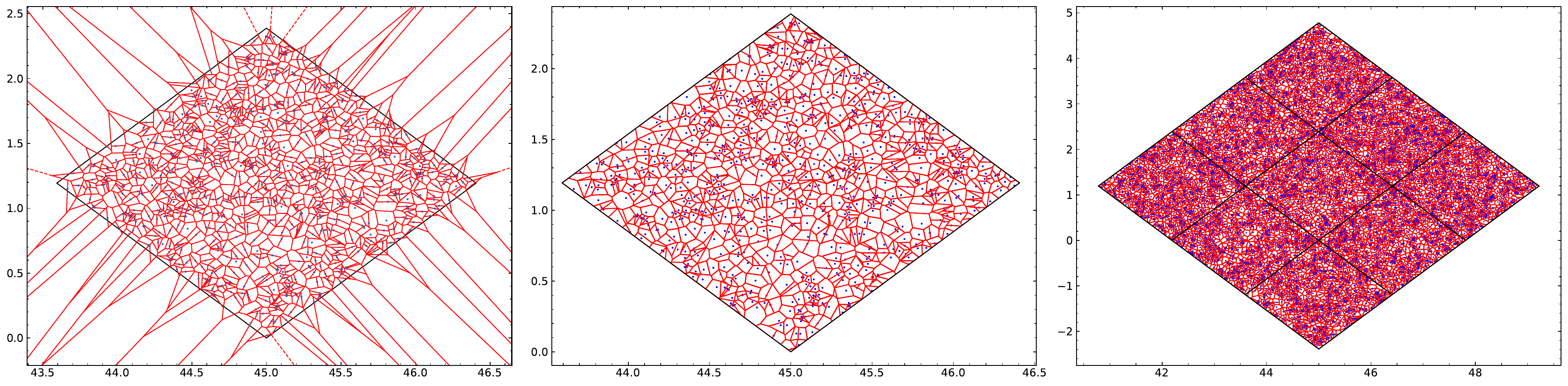}
\caption{Three solutions for the Voronoi algorithm on HEALPix. We have tested these solutions for implementing the Voronoi algorithm on HEALPix. In the left subfigure, we deployed the Voronoi algorithm within a single HEALPix region (PIXArea) but encountered significant edge effects that proved difficult to resolve. We tested the Voronoi algorithm within a PIXArea with a hard boundary, as shown in the middle subfigure. However, we discovered that processing stars in proximity to the PIXArea edge did not yield satisfactory results, as many were erroneously removed due to the abnormally small Voronoi cell near the PIXArea edge. Finally, as depicted in the right subfigure, we adapted the Voronoi algorithm to operate on the center PIXArea and nearby PIXArea. The results obtained within the central PIXArea proved to be reliable and accurate, with no discernible edge effects.}
\label{Sec3_Fig005}
\end{figure}

%--------------------------------------------------------
\begin{figure*}
\centering
\includegraphics[width=0.5\textwidth, angle=0]{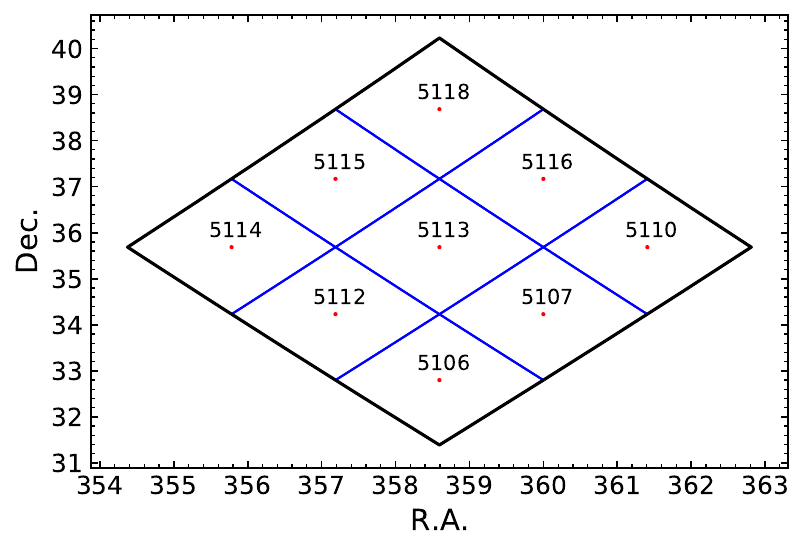}\includegraphics[width=0.5\textwidth, angle=0]{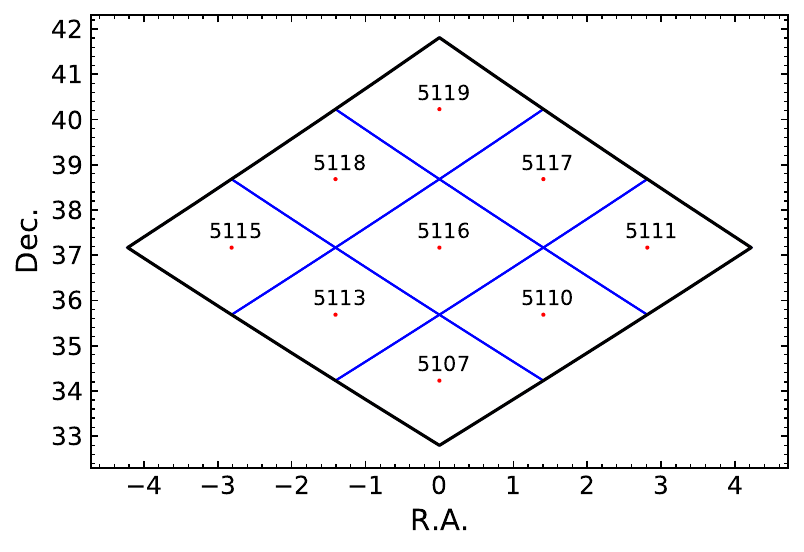}
\caption{The correction of the critical problem (NSIDE = 32). The blue line shows the boundaries of PIXArea, and the black is their joint boundary. The red points are the centers of PIXAreas, and their upper numbers are the HEALPix ID. $left$: most of the centers ($\ge$5) are located at the left of the $0^\circ$ R.A.; $right$: the centers are on the $0^\circ$ R.A. or to its right.} 
\label{Sec3_Fig006}
\end{figure*}

%----------------------------------------------------------
\begin{figure*}
\centering
\includegraphics[width=0.5\textwidth, angle=0, height=0.35\textwidth]{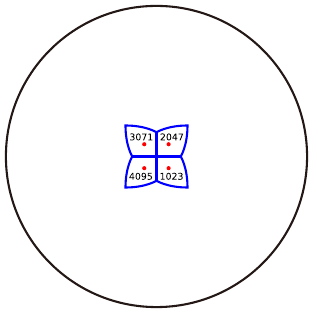}
\includegraphics[width=0.5\textwidth, angle=0]{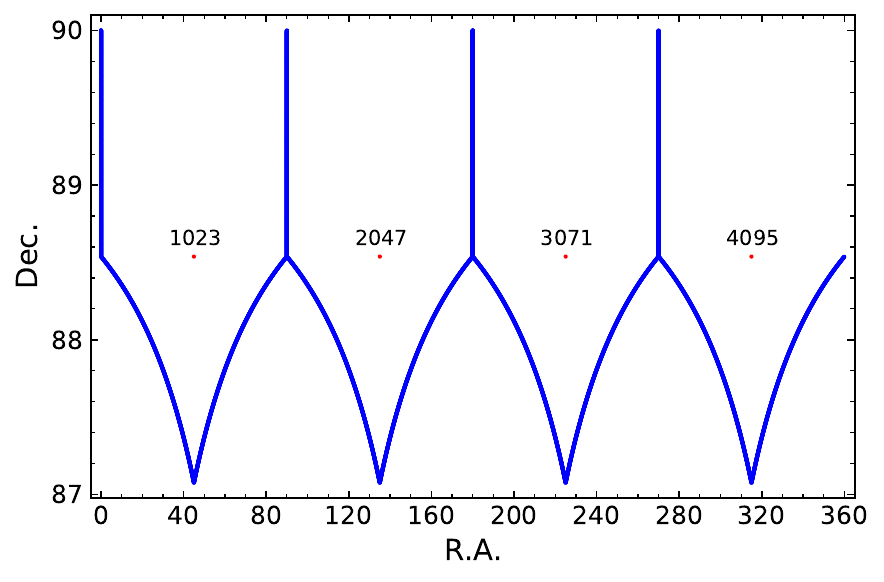}
\caption{The HEALPix region of the North Celestial Poles (NCP). $left$: The picture shows some pixel numbers from the top-down view of the NCP for NSIDE=32 (based on \citealt{2001Splitting}). $right$: The HEALPix regions around the NCP are spread out on the plane for NSIDE=32. }
\label{Sec3_Fig007}
\end{figure*}

%----------------------------------------------------------
\begin{figure*}
\centering
\includegraphics[width=0.5\textwidth, angle=0, height=0.35\textwidth]{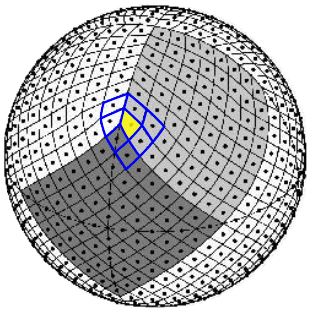}
\includegraphics[width=0.5\textwidth, angle=0]{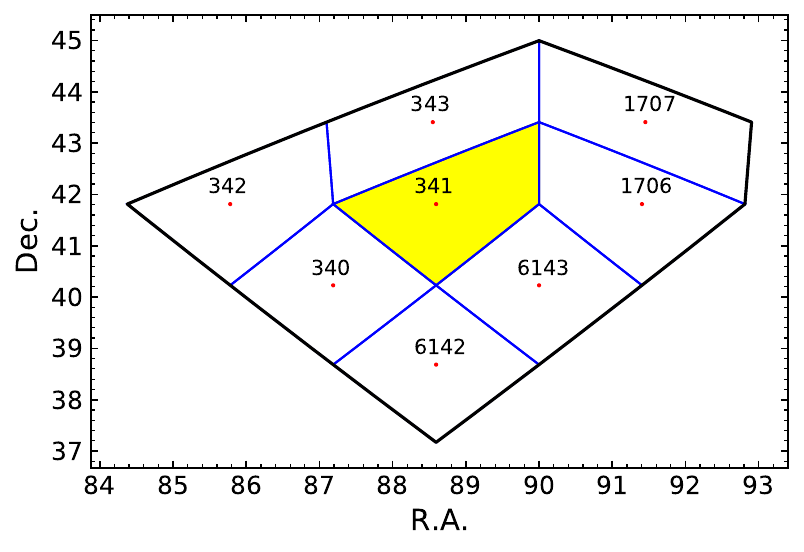}
\caption{The situation for seven near-neighbor regions. $left$: The arrangement of the HEALPix on the sphere (NSIDE = 8). There are three more obvious HEALPix regions at the intersection of white, gray, and black blocks. These three HEALPix regions have seven near-neighbor regions and seven other places on the sphere. $right$: The HEALPix region 341 and its seven neighbors are drawn on the plane (NSIDE = 32).} 
\label{Sec3_Fig008}
\end{figure*}

%--------------------------------------------------------
\begin{figure*}
\centering
\includegraphics[width=0.49\textwidth, angle=0]{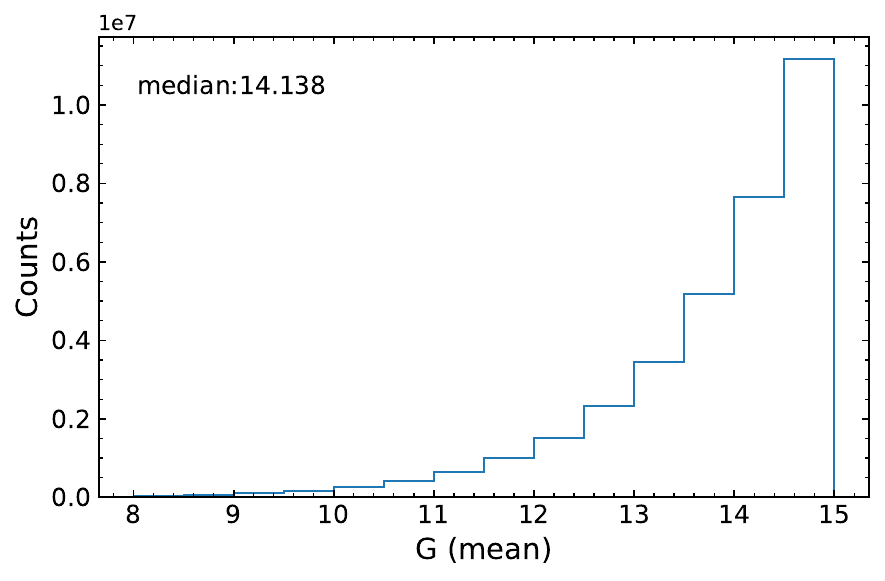}
\includegraphics[width=0.49\textwidth, angle=0]{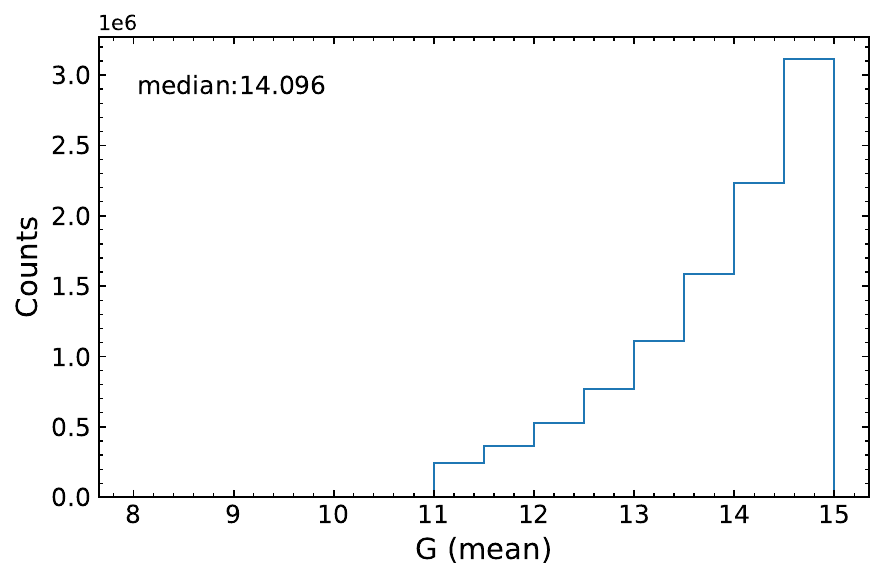}
\caption{$Left$: Magnitude distribution of the preprocessed catalog obtained from the $Gaia$ DR3 main source catalog, after removing binary stars, variable stars, and stars with high proper motion. $Right$: Magnitude distribution of the homogenization results obtained from the preprocessed catalog.} 
\label{Fig006_5}
\end{figure*}

%------------------------------------------------------------
\begin{figure}
\centering
\includegraphics[width=1.0\textwidth, angle=0]{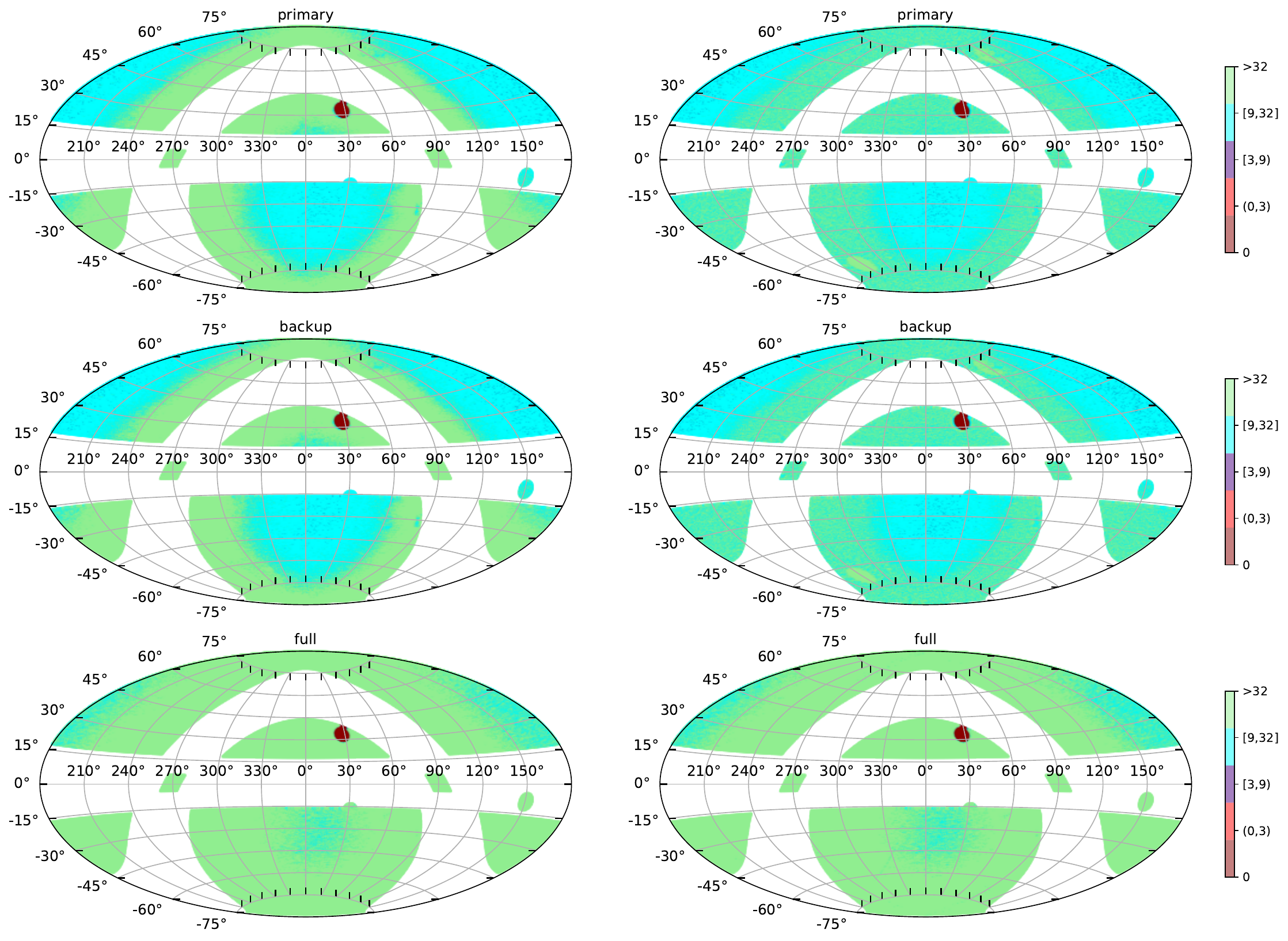}
\caption{The figure illustrates the comparison of guide star number distributions in the ecliptic coordinate system, following the all-sky mock observations, when two MSC FGSs operate in primary, backup, and full operation modes, without (left) and with (right) adapting the homogenization processing. The observations were based on the Mock Survey Strategy Catalog described in Section \ref{subsect:planning}. The color bar represents five bins that indicate the available star number ranges for each observation within the MSC FGSs' FOV. These ranges correspond to the different working conditions of the FGS, including no guide star in FOV (N=0), inability to guide star (0$<$N$<$3), guide star properly (3$\leq$N$<$9), guidance optimized with the selection of the best 9 guide stars (9$\leq$N$\leq$32), and more guide stars than FGS needed (N$>$32). The left column shows the star number distribution in the preprocessed catalog without the homogenization procedure, while the right column shows the distribution with the homogenization procedure applied. The figure is divided into three rows, each corresponding to the MSC FGS operating in primary, backup, and full operation modes.} 
\label{f_1}
\end{figure}

Based on the maximum ROI number and MSC effective area from Table~\ref{Sec2_Tab01}, we can get an ideal star number of MSC in each PIXArea following Eq.~\ref{eq4}. 
\begin{equation}\label{eq4}
   \frac{STAR\_NUMBER}{16\times2} \approx \frac{3.36 deg^{2}}{0.06\times2 deg^{2}}
\end{equation}
where 16 represents the ideal number of the guide stars uniformly distributed in one FGS FOV. 3.36 square degrees is the area of one PIXArea, and 0.06 square degrees is the one FGS effective area in the primary operation mode. There are two FGSs on the MSC. According to Eq.~\ref{eq4}, the Voronoi algorithm implemented within a single PIXArea will stop iterating when the number of stars reduces to 900.

When attempting to homogenize a catalog by combining the HEALPix and Voronoi tessellation methods, several caveats should be addressed:
\begin{itemize}
\item[1.] When using the Voronoi algorithm, it is important to consider the edge effects, which can result in biased estimates of density and other properties near the edges of the survey area. 
As depicted in Figure~\ref{Sec3_Fig005}, we have evaluated various methods of applying the Voronoi algorithm to HEALPix and determined the most optimal solution. When conducting homogenization on a specific PIXArea, it is necessary to apply the Voronoi algorithm to this PIXArea as well as the eight adjacent PIXArea.
\item[2.]In cases where the PIXArea is situated on the great circle that passes through the equinox, north celestial pole, and south celestial pole, the right ascension (R.A.) may undergo significant changes from 360$^\circ$ to 0$^\circ$ degrees. This issue is further exacerbated when considering the eight neighboring PIXAreas while resolving the problem of Voronoi cell crossing. To mitigate this problem, we typically add or subtract 360$^\circ$ from the R.A. of all related PIXAreas. The left side of Figure~\ref{Sec3_Fig006} displays the corrected results for the neighbors of HEALPix ID 5113, while the right side shows the corrected results for the neighbors of HEALPix ID 5116. Alternating between adding and subtracting 360$^\circ$ may be necessary depending on the specific circumstances.
\item[3.]Another issue related to R.A. arises near the celestial pole, where the R.A. varies significantly while the declination (Dec.) remains nearly constant. Besides, as depicted in Figure~\ref{Sec3_Fig007}, the north (or south) celestial pole is not uniquely represented in HEALPix. However, considering the low star density in these PIXAreas, the issue can be addressed by disregarding them. There are a total of 48 PIXAreas remaining that do not require homogenization.
\item[4.]As depicted in Figure~\ref{Sec3_Fig008}, not all HEALPix regions have eight neighboring regions when the NSIDE parameter exceeds 1. Specifically, 24 HEALPix regions only have seven neighboring regions, which require careful handling in the homogenization process.
\item[5.]The Voronoi algorithm is rooted in the two-dimensional plane, whereas celestial coordinates are based on a three-dimensional sphere. Due to the different resolution properties of HEALPix and Voronoi, which can lead to inconsistencies in the resulting catalog, when applying the Voronoi algorithm to celestial coordinates, a significant issue arises known as the "projection effect." However, it is possible to mitigate this effect by increasing the NSIDE parameter, which reduces the size of the HEALPix regions.
\item[6.]The Voronoi algorithm eliminates a single star from the PIXArea during each iteration. Consequently, this algorithm is computationally expensive and time-consuming. To enhance computational efficiency, it is necessary to adapt the parallel processing techniques and optimize the computational workflow while maintaining the accuracy of the sky survey.
\end{itemize}

We can infer from Figure \ref{Fig006_5} that the homogenization procedure decreased star density and total star count, resulting in a more concentrated magnitude distribution. The dynamic range of magnitudes shifted from 8-15 to 11-15, which should be less onerous for the FGS hardware and software to capture reference stars with appropriate signal-to-noise (S/N) ratio.

The FGS deployed on the MSC can operate in three different operation modes: full, primary, or backup. In the full operation mode, all four CMOS sensors are activated. In the primary operation mode, only two primary CMOS sensors are activated, and in the backup operation mode, only two backup CMOS sensors are activated. The MSC FGSs in different modes correspond to the different FOVs. Thus, we compared the sky distribution of the guide star number in the FOVs with and without the homogenization procedure. The results are shown in Figure \ref{f_1} following the ecliptic coordinate. The colors in the figure represent different ranges of guide star numbers. Brown and red indicate that the guide star number is below 3, meaning the FGS cannot function properly in those areas. Purple represents the guide star number between 3 and 9, sufficient for the FGS to operate effectively. Blue and green indicate a guide star number above 9, which allows the MSC FGSs to select the best 9 stars from that range to improve its measurement accuracy and reliability. We could infer from the Figure \ref{f_1} that:

\begin{itemize}
\item[1.] The number of guide stars exhibits some spatial non-uniformity, and the accuracy and stability of FGS operation depends on the sky regions. This spatial non-uniformity distribution is correlated strongly with galactic latitude.

\item[2.] Comparing the first, second, and third rows of Figure \ref{f_1}, it can be inferred that there is no change in the FGS performance in the primary and backup operation modes. However, operating in the full operation mode significantly improves the number of guide stars in some sky regions where the guide star count is initially low. This indicates that utilizing the full operation mode can potentially enhance the FGS's performance in these regions. However, it is important to note that operating the FGS in the full operation mode will result in a marked increase in power consumption. Thus, further research and planning may be needed to optimize the CSST sky survey strategy and FGS operation mode in these areas.

\item[3.] Comparing the left and right columns of Figure \ref{f_1}, it can be seen that the implementation of the homogenization procedure does not have a significant impact on the spatial distribution of the number of guide stars. Furthermore, it will not affect the FGS operation performance across the catalog based on the results and verification provided in Table \ref{Tab}. 

\item[4.] Due to the variable stars being removed from the catalog, all stars in the M31 region present in the $Gaia$ DR3 variable star catalog are missing. Therefore, a recall is necessary to call back the stars in the M31 region.

\end{itemize}

Table \ref{Tab} presents a summary of the number of MSC FGSs captured guide stars in various ranges ((0), (0, 3), [3, 9), [9, 32], and (N$>$32)) in primary, backup, and full operation modes. The table also includes whether catalog homogenization processing was applied during all-sky mock observations, as depicted in Figure~\ref{f_1}. Moreover, we have transformed it into Table \ref{Tab_comb} to assess the impact of catalog homogenization processing by evaluating the probabilities of FGS operation conditions. From the table, we can infer that catalog homogenization processing does not significantly affect the probabilities of the FGS operating in the optimized condition when it is in full operation mode. However, catalog homogenization processing slightly decreases the probabilities by about 0.1\% when the FGS is in primary or backup operation mode. The difference of probabilities in the condition of FGS unable guide star, with or without catalog homogenization processing, is about 0.01\%. In summary, adopting the homogenization procedure has a small impact on the FGS operation conditions. Given the benefits of significantly reducing storage space requirements and data computation complexity in orbit, homogenization processing is necessary for the CSST guide star catalog.

%________________________________________ 
\begin{table} \centering
\begin{minipage}[]{\textwidth}
\caption{Guide star number distribution in primary, backup, and full operation modes of MSC FGSs with and without homogenization processing in all-sky mock observations, corresponding to Figure~\ref{f_1}. 
\label{Tab}} \end{minipage} \\
\setlength{\tabcolsep}{2mm}
\small
 \begin{tabular}{cccccccc}
  \hline
  \hline\noalign{\smallskip}
Homogenization & FGS Mode & $N$=0 &	$N\in$ (0, 3) &	$N\in$ [3, 9)  &	$N\in$ [9, 32] &	N $>$ 32  & Total Observations  \\
  \hline
  \hline\noalign{\smallskip}
\multirow{3}{*}{Without}& primary 	& 2295 & 40 & 1020 & 257 860 & 355 043   & \multirow{3}{*}{616 258} \\
                     & backup   & 2260 & 37 & 1004 & 257 469 & 355 488	& ~ \\
                     & full	   & 2190 & 55 & 37 & 33 478 & 580 498	& ~  \\ \hline
\multirow{3}{*}{With}& primary	& 2304 & 44 & 1797 & 380 749 & 231 364	& \multirow{3}{*}{616 258} \\
                        & backup	& 2268 & 39 & 1751 & 380 175 & 232 025	& ~  \\
                        & full	    & 2202 & 49 & 43 & 39 576 & 574 388	& ~  \\ 
  \noalign{\smallskip}\hline
  \hline
\end{tabular}
\tablecomments{\textwidth}{"Without" and "With" Homogenization correspond to the left and right columns in Figure~\ref{f_1}. "Total Observations" is the total number of mock observations following the Mock Survey Strategy Catalog.}
\end{table}

%________________________________________ 
\begin{table} \centering
\begin{minipage}[]{\textwidth}
\caption{The probabilities of MSC FGSs' operation conditions following the Mock Survey Strategy Catalog, derived from the Table \ref{Tab}.
\label{Tab_comb}} \end{minipage} \\
\setlength{\tabcolsep}{3.5mm}
\small
 \begin{tabular}{cccccccc}
  \hline
  \hline\noalign{\smallskip}
Homogenization & FGS Mode &	$N\in$ [0, 3) &	$N\in$ [3, 9)  &	$N\in$ [9, $\infty$) \\
  \hline
  \hline\noalign{\smallskip}
\multirow{3}{*}{Without}& primary	& 0.38\% & 0.17\% & 99.46\% \\
                     & backup   & 0.37\% & 0.16\% & 99.46\% \\
                     & full	   & 0.36\% & 0.0\% & 99.63\%	\\ \hline
\multirow{3}{*}{With}& primary	& 0.39\% & 0.29\% & 99.33\%	\\
                        & backup	& 0.37\% & 0.28\% & 99.34\%	\\
                        & full	    & 0.37\% & 0.0\% & 99.63\% \\ 
  \noalign{\smallskip}\hline
  \hline
\end{tabular}
\tablecomments{\textwidth}{ The FGS operation conditions defined as $N\in$ [0, 3): FGS unable to guide star; $N\in$ [3, 9): FGS able to guide star; $N\in$ [9, $\infty$) : FGS works in optimized conditions.
}
\end{table}

\subsection{The Catalog Performance} 
\label{subsect:all_sky}
It is evident from Figure \ref{f_1} that the density of guide stars is uneven across the entire sky, particularly in regions with high galactic altitudes where the guide star density is relatively low. Additionally, there is a region in M31 where guide stars are absent due to the removal of the variable stars. Thus, the reserve of the GAPS in M31 is necessary, although 1\% variables are mixed in it. We are curious if this method could increase the guide star number and improve the FGS performance in other sky areas, even if the accuracy of the catalog decreased slightly. In addition, We can also infer from Table \ref{Tab} and Table \ref{Tab_comb} that operating the FGS in full operation mode results in an increased number of FGS-captured guide stars. 
However, this mode can result in increased power consumption and reduced lifespan of the FGS, potentially impacting the observation quality of the CSST in later stages in space. 
Determining whether to activate all sensors in the FGS involves a trade-off between performance and other adverse conditions mentioned above. Hence, conducting a preliminary test before completing the Main Guide Star Catalog is advisable. This test should assess the operating conditions of the FGS in primary, backup, and full operation modes across various sky areas, considering the presence of variable stars.

We have initially selected three sky regions, each covering an area of 10 square degrees, in low, medium, and high galactic latitude. The celestial centers of these regions, given in galactic coordinates, are (185.452$^{\circ}$, 0.507$^{\circ}$), (279.308$^{\circ}$, -50.780$^{\circ}$), and (100.470$^{\circ}$, 89.999$^{\circ}$). These regions correspond to a total of 312, 310, and 312 observations of MSC in the Mock Survey Strategy Catalog, respectively. Additionally, we have retrieved and processed all stars in GAPS, then selected 2289 mock observations of MSC in the M31 region. Furthermore, we have included 192 mock observations of IFS for comparison, even though FGS on SFP can only operate in full operation mode. In these selected regions, there are a total of 853 PIXAreas that are pending and form a sub-catalog. 
This sub-catalog has undergone preprocessing and homogenization processing, involving variable stars or not. The gnomonic projection method is adapted to determine the FGS-captured guide stars, referring to Section~\ref{subsect:GnoPro}. Based on the two sets of guide stars, we have obtained the results presented in Table~\ref{table6_new}.

Based on the information provided in Table~\ref{table6_new}, it can be inferred that the number of guide stars in low and medium galactic latitude areas is sufficient for the MSC FGSs to operate optimally (N$\in$ [9, $\infty$)) in both primary and backup modes. Therefore, activating all four sensors is unnecessary in these areas. However, in high galactic latitude areas, activating all four sensors of the MSC FGSs can result in a slight improvement in its operation, transitioning from proper functioning (N$\in$ [3, 9)) to optimal performance (N$\in$ [9, $\infty$)). This test demonstrates an approximate 4\% improvement in performance based on the results presented in Table~\ref{table6_new}. Furthermore, Table~\ref{table6_new} illustrates that including variable stars in the guide star catalog resolves the issue of guide star unavailability in the M31 region. It also highlights the slight improvement in FGS performance in high galactic latitude areas and resolves the issue of disabled IFS observation.

To ensure optimal performance and minimize overhead, it is important to carefully manage the frequency of mode conversion for the MSC FGS by taking into account the lengthy cool-down and self-check periods associated with activating additional sensors. In high galactic latitude regions, it is advisable to operate the MSC FGSs in full operation mode to maximize the number of available guide stars. Variable stars should be specifically reserved for observations in the M31 region. Furthermore, it is recommended to allocate the variable stars for IFS observations as this meets the more stringent guide star requirements resulting from multiple exposures achieved by rotating around a center. The primary version of the guide star catalog has considered the findings above. However, our study suggests that further analysis and research are necessary to optimize the survey strategy based on the distribution of guide stars.

\begin{figure*}
\centering
\includegraphics[width=0.49\textwidth, angle=0]{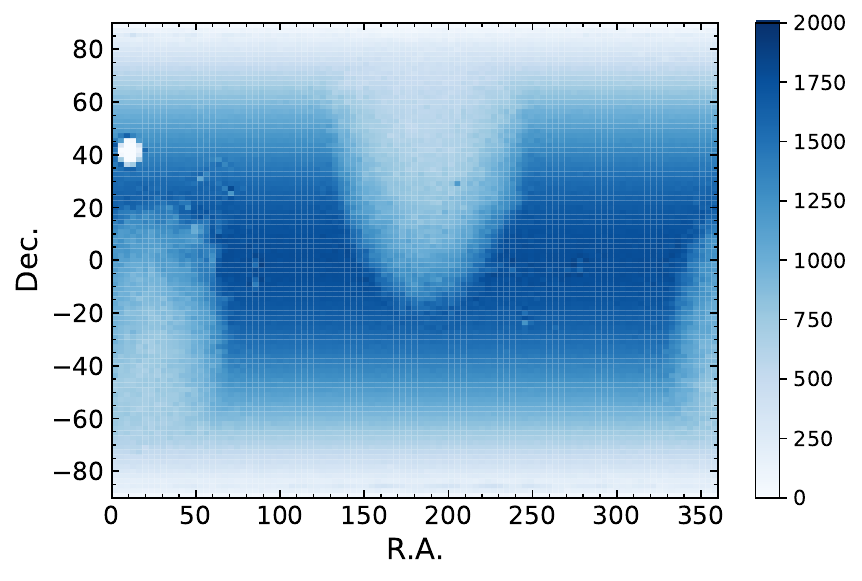}
\includegraphics[width=0.49\textwidth, angle=0]{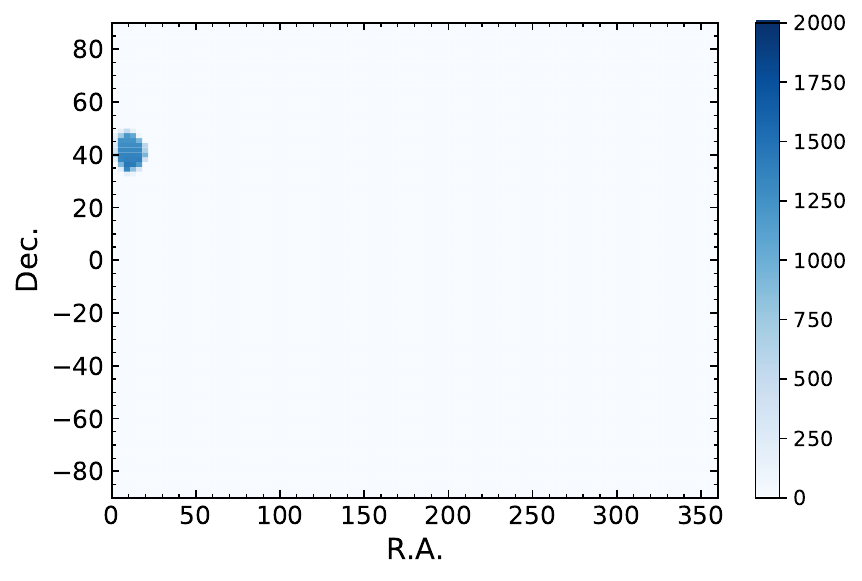}
\caption{$left$: The density distribution of all-sky homogenization results for G=11-15. $right$: The density distribution of supplementary homogenization results of M31.} 
\label{f_4}
\end{figure*}

\begin{figure*}
\centering
\includegraphics[width=0.44\textwidth, height=0.3\textwidth]{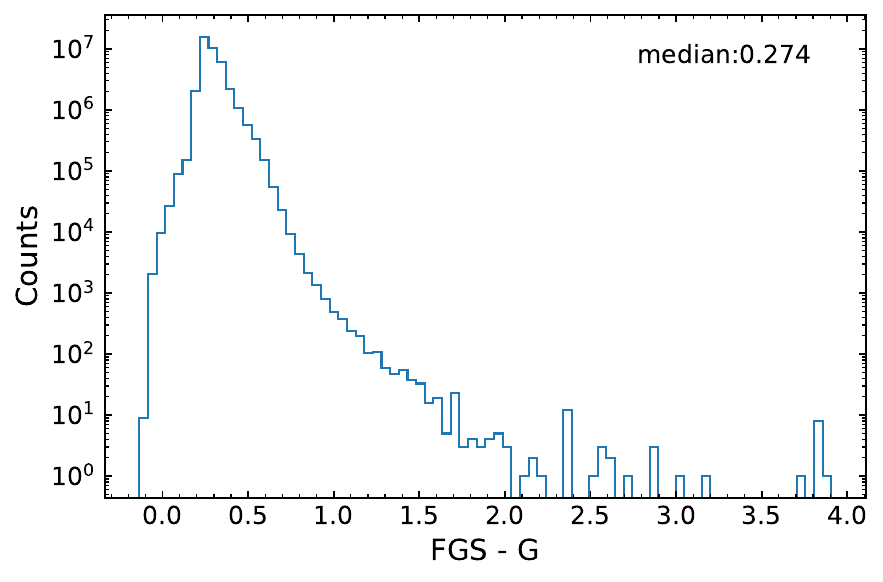}
\includegraphics[width=0.55\textwidth, height=0.3\textwidth]{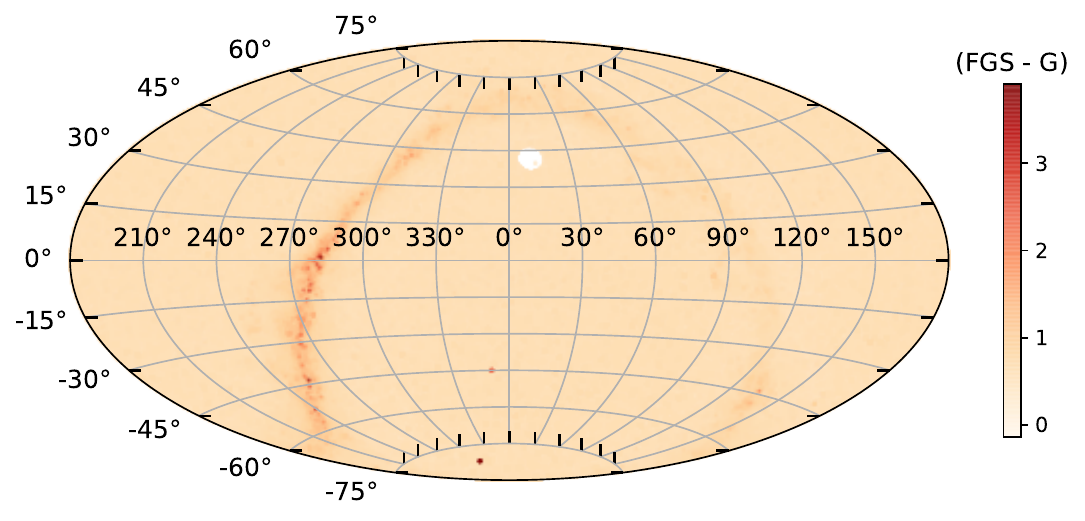}
\caption{The distribution of expected instrument magnitude ($FGS$) minus $Gaia$ $G$. $left$: the histogram of $(FGS-G)$. $right$: the spatial distribution of $(FGS-G)$ in equatorial coordinate system.} 
\label{f_5}
\end{figure*}

Figure \ref{f_4} illustrates the density distribution of stars in CSST-MGSC, with the left sub-figure representing the entire sky excluding M31 and the right sub-figure focusing specifically on the M31 region. Compared to Figure \ref{Sec5_gaiadr3_nobnov}, the density of stars appears to be more evenly distributed across the sky. Furthermore, the star number density in M31, known for its crowded-field feature, is consistent with its surrounding areas. We also present the histogram and spatial distribution of the derived instrument magnitude (FGS) minus $Gaia$ $G$ magnitude in Figure \ref{f_5}. In the left sub-figure, the median value of (FGS-G) is 0.274, with several high values exceeding 2.0. The right sub-figure reveals a uniform distribution of (FGS-G) across the entire sky, except for elevated values along the galactic plane towards the galactic center and relatively higher values along the galactic plane towards the anti-galactic center direction. We can deduce that these high values of (FGS) are primarily due to the extremely high extinction and significant reddening in the Milky Way disk. However, it is possible that system errors of the $LAMOST$ template spectrum partly influence this distribution in these specific regions.

We have verified that the catalog followed the method described in Section~\ref{subsect:GnoPro} and matched the guide stars from the guide star catalog with each mock observation. There are 616,258 different MSC observation centers in the Mock Survey Strategy Catalog. To facilitate the analysis, we divided the mock observation centers into five groups based on galactic latitude. These groups are categorized as follows: low galactic latitude ((-45$^{\circ}$, 45$^{\circ}$)), medium galactic latitude ((-75$^{\circ}$, -45$^{\circ}$], [45$^{\circ}$, 75$^{\circ}$)), and high galactic latitude ([-90$^{\circ}$, -75$^{\circ}$], [75$^{\circ}$, 90$^{\circ}$]). Specifically, within the low galactic latitude region, there are 377,270 observation centers. In the medium galactic latitude regions, there are 197,602 observation centers. Lastly, the high galactic latitude regions have 41,386 observation centers.
We configured the MSC FGS to operate in the primary operation mode for verification purposes in low and medium galactic latitude regions. In high galactic latitude regions, we set the MSC FGS to operate in full operation mode and reserve primary operation mode for comparison. To ensure fully optimized guidance, we required the MSC FGSs to capture a minimum of 15 stars. From these captured stars, we iteratively selected 9 best quality guide stars for guidance. The "Captured probability" and "Guide probability" represent the ratios of the observations that meet these requirements to the total number of observations in each sky region. The "MGSC Accuracy" is determined based on the astrometry accuracy of the guide star catalog. It is calculated by taking the median of the vector sums of the "ra\_err" and "dec\_err" values of the matched guide stars in each sky area.
According to CSST engineering requirements, the FGS "Guide Probability" should be above 95\% in MSC observations. The verification results are presented in Table~\ref{Tab200}. The "Guide Probability" exhibits similar values in low and medium galactic latitude regions. However, compared to low galactic latitude regions, the "Captured Probability" is relatively low at approximately 10.5\%, due to the lower star density in medium galactic latitude. The statistics differ in high galactic latitude regions. The "Captured Probability" drops to 66.86\%, and the corresponding "Guide Probability" reaches its lowest value of 97.97\%. When the MSC FGS operates in its full operation mode, these statistics improve to 99.96\% and 100\%, respectively. The "MGSC Accuracy" in different regions is nearly identical, with a value of around 0.02 mas. 

%------------------------------------------------------------
\begin{table}[t]
\setlength{\tabcolsep}{1mm}
    \caption{Guide star number distribution in the primary, backup, and full operation modes of the MSC FGSs varies across different observational scenarios. The first column abbreviates the region of the sky considered as low, medium, high, M31, and SFP representing low galactic latitude MSC mock observations, medium galactic latitude MSC mock observations, high galactic latitude MSC mock observations, M31 MSC mock observations, and all-sky mock observations respectively. Variables are removed in the left table, and variables are retained in the right table (including all stars in GAPS).} 
    \label{table6_new}
    \begin{minipage}{0.5\textwidth}
      \centering
        (a) Probability (without variables)
        \resizebox{7.5cm}{3.25cm}{
        \small
        \begin{tabular}{cccccc}
            \hline
  \hline\noalign{\smallskip}
Region & Mode & $N \in$ [0, 3) &	$N \in$ [3, 9)  &	$N \in$ [9, 32]  &	 $N >$ 32    \\
  \hline
  \hline\noalign{\smallskip}
\multirow{3}{*}{Low}& primary	& 0\% & 0\%	& 38.78\% & 61.22\%	 \\
                    & backup	& 0\% & 0\%	& 40.06\% & 59.94\%	 \\
                    &  full	    & 0\% & 0\%	& 0\%	    & 100\%	 \\ \hline
\multirow{3}{*}{Medium}& primary	& 0\% & 0\%	& 78.39\% & 21.61\%	 \\
                       & backup	& 0\% & 0\%	& 76.77\% & 23.23\%	 \\
                       &  full	    & 0\% & 0\%	& 0\%	    & 100\%	 \\ \hline
\multirow{3}{*}{High}& primary	& 0\% & 4.17\% & 95.83\%	& 0\%	 \\
                     & backup	& 0\% & 3.85\% & 96.15\%	& 0\%	 \\
                     & full	    & 0\% & 0\%	& 57.37\% & 42.63\%	 \\ \hline
\multirow{3}{*}{M31}& primary	& 98.47\% & 1.53\%	& 0\% & 0\%	 \\
                    & backup	& 97.68\% & 2.18\%	& 0.13\% & 0\%	 \\
                    & full	    & 96.2\% & 3.63\%	& 0.17\%	    & 0\%	 \\ \hline
SFP & full	    & 0.52\% & 3.13\% & 95.83\% &	0.52\%	 \\
  \noalign{\smallskip}\hline
  \hline
  \end{tabular}
    }
    \end{minipage}
    \begin{minipage}{0.5\textwidth}
     \centering
        (b) Probability (with variables)
        \resizebox{7.5cm}{3.25cm}{
        \begin{tabular}{cccccc}
            \hline
  \hline\noalign{\smallskip}
Region & Mode & $N \in$ [0, 3) &	$N \in$ [3, 9)  &	$N \in$ [9, 32]  &	 $N >$ 32    \\
  \hline
  \hline\noalign{\smallskip}
\multirow{3}{*}{Low}& primary	& 0\% & 0\%	& 38.78\% & 61.22\%	 \\
                    & backup	& 0\% & 0\%	& 40.06\% & 59.94\%	 \\
                    &  full	    & 0\% & 0\%	& 0\%	    & 100\%	 \\ \hline
\multirow{3}{*}{Medium}& primary	& 0\% & 0\%	& 78.39\% & 21.61\%	 \\
                       & backup	& 0\% & 0\%	& 76.77\% & 23.23\%	 \\
                       &  full	    & 0\% & 0\%	& 0\%	    & 100\%	 \\ \hline
\multirow{3}{*}{High}& primary	& 0\% & 2.88\% & 97.12\%	& 0\%	 \\
                     & backup	& 0\% & 1.92\% & 98.08\%	& 0\%	 \\
                     & full	    & 0\% & 0\%	& 46.15\% & 54.17\%	 \\ \hline
\multirow{3}{*}{M31}& primary	& 0\% & 0\%	& 37.83\% & 62.17\%	 \\
                    & backup	& 0\% & 0\%	& 39.45\% & 60.55\%	 \\
                    & full	    & 0\% & 0\%	& 0\%	    & 100\%	 \\ \hline
SFP & full	    & 0\% & 2.6\% & 96.88\% &	0.52\%	 \\
  \noalign{\smallskip}\hline
  \hline
  \end{tabular}
  }
  \end{minipage} 

\end{table}

%--------------------------------------------------------

%--------------------------------------------------------

%________________________________________ 
\begin{table} \centering
\begin{minipage}[]{\textwidth}
\caption{Verification of catalog performance through the mock survey strategy.
\label{Tab200}} \end{minipage} \\
\setlength{\tabcolsep}{1mm}
\small
 \begin{tabular}{ccccc}
  \hline
  \hline\noalign{\smallskip}
 No. & Items & Low Galactic Latitude & Medium Galactic Latitude & High Galactic Latitude\\
    &   & (-45$^{\circ}$, 45$^{\circ}$) & (-75$^{\circ}$, -45$^{\circ}$] or [45$^{\circ}$, 75$^{\circ}$) & [-90$^{\circ}$, -75$^{\circ}$] or [75$^{\circ}$, 90$^{\circ}$]\\
  \hline\noalign{\smallskip}
1 & Mean Magnitude (${M_v}$) & 13.75 & 13.66 & 13.64 \\
2 & Smallest Angular distance (ROI) & 0.6$^{''}$ & 0.6$^{''}$ & 0.6$^{''}$ \\
3 & Captured Star Number & $\ge$ 15 & $\ge$ 15 & $\ge$ 15 \\
4 & Captured Probability (\%) & 99.88 & 89.38 & 99.96 / 66.86 \\
5 & Calculated Star Number & 9 & 9 & 9 \\
6 & Guide Probability (\%) & 99.98 & 99.54 & 100.00 / 97.97\\ % all >=9 stars
7 & MGSC Accuracy (mas) & 0.02 & 0.02 & 0.02 / 0.02 \\
8 & Mode & Primary & Primary & Full / Primary\\
  \noalign{\smallskip}\hline
  \hline
\end{tabular}
\end{table}

\section{The Catalog Release} \label{sect:mgsc_release}
This catalog is available in the China-VO PaperData Repository, provided by China National Astronomical Data Center (NADC), CAS Astronomical Data Center, and Chinese Virtual Observatory (China-VO) \footnote{https://nadc.china-vo.org/}. Table~\ref{Tab001} presents the metadata of this catalog. The "phot\_inst\_mean\_mag" is the theoretical FGS instrument magnitude derived from the $Gaia$ G, BP, and RP mean magnitude. The reduction is based on the $LAMOST$ template spectrum (\citealt{2014AJ.147.101W}), and the processing will be described in Section~\ref{subsect:instmag}. The residual of this processing is recorded in "phot\_inst\_mean\_mag\_error". The source quality flag ("source\_flag") in Table~\ref{Tab001} indicates the activation status of the backup sensors. It can have three values: "primary", "backup", and "full", which stand for the primary operation mode, backup operation mode, and full operation mode, respectively. The mock survey strategy catalog comprises all the information regarding CSST observations throughout its 10-year orbital mission. Based on the mock survey strategy catalog, the verification process described in Section~\ref{subsect:GnoPro} determines whether the guide star is within the field of view of the FGS. If a star falls within the MSC FGS FOV, the "msc\_skyarea\_id" field indicates the corresponding sky area ID from the mock survey strategy catalog, and the "msc\_cmos\_id" field indicates the CMOS ID of the star's location. Otherwise, these fields will be empty. Similarly, if a star is within the SFP FGS FOV, the "sfp\_skyarea\_id" field indicates the corresponding sky area ID from the mock survey strategy catalog, and the "sfp\_cmos\_id" field indicates the CMOS ID of the star's location. Otherwise, these fields will be empty.

%________________________________________
\begin{table}\centering
\begin{minipage}[]{\textwidth}
\caption{Metadata of the Guide Star Catalog (CSST-MGSC).
\label{Tab001}} \end{minipage} \\
\setlength{\tabcolsep}{5mm}
\small
 \begin{tabular}{ccc}
  \hline
  \hline\noalign{\smallskip}
 Column &   Name  & Description   \\
  \hline
  \hline\noalign{\smallskip}
1  & source\_id        &   $Gaia$ DR3 source ID      \\ 
2  & guide\_ra         &   $Gaia$ Right ascension (ICRS) at epoch 2016.0      \\
3  & guide\_ra\_error   &    Standard error of right ascension      \\
4  & guide\_dec        &   $Gaia$ Declination (ICRS) at epoch 2016.0      \\
5  & guide\_dec\_error  &    Standard error of declination      \\
6  & pm     &    Total proper motion      \\
7  & pmra             &   $Gaia$ DR3 proper motion in $R.A. \times $cos $DEC$      \\
8  & pmra\_error   &   Standard error of proper motion in right ascension direction       \\
9  & pmdec            &   Proper motion in $DEC$      \\
10 & pmdec\_error  &   Standard error of proper motion in declination direction       \\
11 & phot\_g\_mean\_mag  &   $Gaia$ $G$ magnitude      \\
12 & phot\_g\_mean\_flux   &   $Gaia$ $G$ band mean flux       \\
13 & phot\_g\_mean\_flux\_error  &  Error on $G$-band mean flux        \\
14 & ruwe                 &    Renormalised unit weight error      \\
15 & ecl\_lon             &   Ecliptic longitude        \\
16 & ecl\_lat             &   Ecliptic latitude     \\
17 & bp\_rp   &   BP-RP colour       \\
18 & phot\_inst\_mean\_mag  &   Instrument magnitude      \\
19 & phot\_inst\_mean\_mag\_error  &   Instrument magnitude error      \\
20 & variable\_flag  &   Is a variable star, or not \\
21 & source\_flag  &   Indicates the activation status of the backup sensors      \\
22 & msc\_skyarea\_id      &   Sky area id from the mock survey strategy catalog of MSC      \\
23 & msc\_id\_cmos      &   CMOS id of MSC which source locates      \\
24 & sfp\_skyarea\_id      &   Sky area id from the mock survey strategy catalog of SFP      \\
25 & sfp\_id\_cmos      &   CMOS id of SFP which source locates      \\

  \noalign{\smallskip}\hline
  \hline
\end{tabular}
\end{table}

\section{Conclusion}
\label{sect:summary}
We have built an all-sky guide star catalog for CSST, named CSST-MGSC. This catalog takes advantage of the $Gaia$ DR3 Catalog's excellent sky uniformity and high astrometric accuracy. To ensure the quality of the CSST-MGSC, we have excluded variables, binaries, and high proper motion stars. We have employed HEALPix and Voronoi algorithms to achieve homogeneity within the catalog. These algorithms generate homogenized sub-area catalogs and limit the magnitude range to 11-15 mag. Through subsequent analysis, we have determined that the Voronoi algorithm is essential in constructing the CSST-MGSC, although it may somewhat reduce star density. Additionally, we have derived theoretical instrument magnitudes based on the $G$, $BP$, $RP$, and FGS system transmission data. 

Before completing the catalog, we chose four testing regions for the MSC and conducted 192 IFS observations for the SFP. To verify the catalog, we have employed the gnomonic projection method. The test results indicate that variable stars should be retained in the M31 region. Preserving them in high galactic latitude regions and IFS observation regions is beneficial to increase the availability of guide stars. Additionally, activating all FGS sensors in high galactic latitude regions is suggested as it can improve the performance of the FGS. These findings provide valuable insights into optimizing the catalog and enhancing the guide star selection process.

After completing the processing of the CSST-MGSC, we conducted an all-sky verification of the catalog based on different galactic latitudes. While the statistics vary across different regions, the "Guide Probability", which represents the probability of capturing guide stars with a number larger than 9 by the FGS, exceeds 95\%. The verification process demonstrates that the CSST-MGSC is highly reliable and can provide sufficient guide stars for the CSST FGS. The catalog's performance meets the FGS requirements, ensuring the proper functioning of the FGS and its guidance capabilities in the CSST sky survey.

\appendix
\section{The Gnomonic Projection} \label{sec:GnoPro}
\begin{figure}
\centering
\includegraphics[width=\textwidth, angle=0]{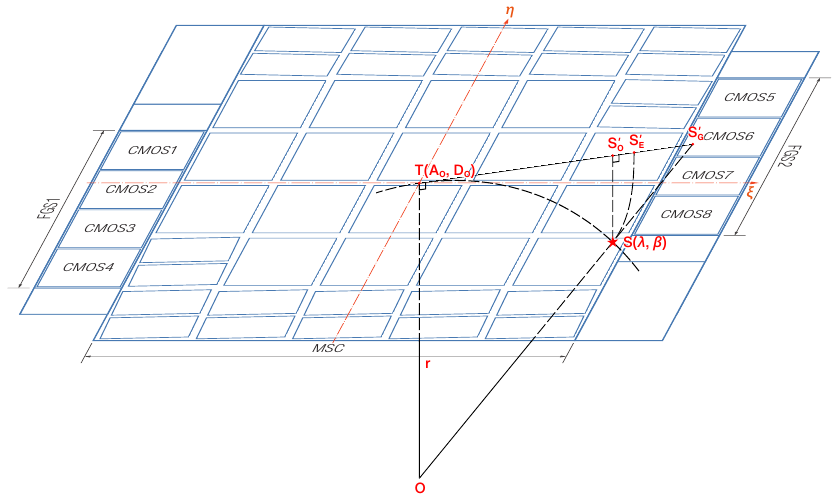}
%\captionsetup{justification=raggedright,width=\linewidth,skip=5pt}
\caption{The gnomonic projections of the celestial sphere onto the MSC. The focal plane is tangent to the celestial sphere at $T$. The $\eta$ axis targets the projection of the ecliptic pole along the ecliptic longitudes, and the $\xi$ axis is parallel to the ecliptic latitude and increasing by the direction of spherical ecliptic longitude.}
\label{Sec3_Fig009}
\end{figure}

Taken FGSs on MSC, for example. There are three classical modes to project the point on a sphere onto a tangent plane to construct the standard coordinate \citep{Dick1991, richardos1972map}, including the gnomonic projection ($S^\prime_{G}$ on Figure~\ref{Sec3_Fig009}), the azimuthal equidistant projection ($S^\prime_{E}$), and the orthographic projection ($S^\prime_{O}$). 
Based on the gnomonic projection, we introduce a rectangular system of standard coordinate $\xi$ and $\eta$ (Figure~\ref{Sec3_Fig009}) in the focal plane, which is origins in the center $T(A_O, D_O)$ in ecliptic coordinate, corresponding to the standard coordinate origin $O(0, 0)$. 
There is a strict transformation conversion (\citealt{Dick1991}) from ecliptic coordinates $(\lambda, \beta)$ to standard coordinates $(\xi, \eta)$ of the stars, following: 
\begin{equation}\label{eq1}
   \xi=\frac{\cos\beta\cdot~\sin(\lambda-A_O)}{\sin\beta\cdot~\sin D_O + \cos\beta\cdot~\cos D_O\cdot ~\cos (\lambda - A_O)}
\end{equation}
\begin{equation}\label{eq2}
   \eta=\frac{\sin\beta\cdot~\cos D_O-\cos\beta\cdot~\sin D_O\cdot~\cos(\lambda-A_O)}{\sin\beta\cdot~\sin D_O + \cos\beta\cdot~\cos D_O \cdot~\cos (\lambda - A_O)}
\end{equation}
where $(A_O, D_O)$ is the ecliptic coordinate of telescope pointing; $(\lambda, \beta)$ is the ecliptic coordinate of a star; $(\xi, \eta)$ is the corresponding standard coordinate of this star.

%------------------------------------------------------------

\section{The Sensors' Positions} \label{sec:sensor_coord} 
Each MSC FGS involves four CMOSs. Table \ref{Sec3_Tab002} details their geometric positions in standard coordinates. Additionally, each MSC FGS involves two CMOSs, and their geometric positions in standard coordinates are in Table \ref{Sec3_Tab003}. It is essential to highlight that the selection of guide stars for each sensor should consider a location range slightly smaller than the sensor's FOV, which is contingent upon the dimensions of the FGS ROI. 
Furthermore, the existing geometric positions of each sensor are based on the initial FGS design specifications. For enhanced accuracy, the sensor geometric positions should be conducted using laboratory and on-orbit calibration methodologies, including the static aberrations inherent to the optics and the dynamic distortions arising from active optics. 

%________________________________________ 
\begin{table}\centering
\begin{minipage}[]{\textwidth}
\caption{The standard coordinates of the eight CMOSs' geometric corners on the MSC, corresponding to Figure~\ref{Sec2_Fig002}.
\label{Sec3_Tab002}} \end{minipage} \\
\setlength{\tabcolsep}{3.5mm}
\small
 \begin{tabular}{ccccccccc}
  \hline
  \hline\noalign{\smallskip}
No. &\multicolumn{2}{c}{Left Top} & \multicolumn{2}{c}{Left Down} & \multicolumn{2}{c}{Right Down} &\multicolumn{2}{c}{Right Top} \\
&  $\xi$  & $\eta$ &  $\xi$  & $\eta$ &  $\xi$  & $\eta$&  $\xi$  & $\eta$ \\ 
  \hline
  \hline\noalign{\smallskip}
CMOS1 & -0.75 & 0.1568 & -0.75 & 0.0389 & -0.5817 & 0.0389 & -0.5817 & 0.1568 \\
CMOS2 & -0.75 & 0.0125 & -0.75 & -0.1054 & -0.5817 & -0.1054 & -0.5817 & 0.0125 \\
CMOS3 & -0.75 & -0.1318 & -0.75 & -0.2496 & -0.5817 & -0.2496 & -0.5817 & -0.1318 \\
CMOS4 & -0.75 & -0.276 & -0.75 & -0.3939 & -0.5817 & -0.3939 & -0.5817 & -0.276 \\
CMOS5 & 0.5817 & 0.3939 & 0.5817 & 0.276 & 0.75 & 0.276 & 0.75 & 0.3939 \\
CMOS6 & 0.5817 & 0.2496 & 0.5817 & 0.1318 & 0.75 & 0.1318 & 0.75 & 0.2496 \\
CMOS7 & 0.5817 & 0.1054 & 0.5817 & -0.0125 & 0.75 & -0.0125 & 0.75 & 0.1054 \\
CMOS8 & 0.5817 & -0.0389 & 0.5817 & -0.1568 & 0.75 & -0.1568 & 0.75 & -0.0389 \\
  \noalign{\smallskip}\hline
  \hline
\end{tabular}
\end{table}

%________________________________________ 
\begin{table}\centering
\begin{minipage}[]{\textwidth}
\caption{The standard coordinates of the four CMOSs' geometric corners on the SFP, corresponding to Figure~\ref{Sec2_Fig003}.
\label{Sec3_Tab003}} \end{minipage} \\
\setlength{\tabcolsep}{3.5mm}
\small
 \begin{tabular}{ccccccccc}
  \hline
  \hline\noalign{\smallskip}
No. &\multicolumn{2}{c}{Left Top} & \multicolumn{2}{c}{Left Down} & \multicolumn{2}{c}{Right Down} &\multicolumn{2}{c}{Right Top} \\
&  $\xi$  & $\eta$ &  $\xi$  & $\eta$ &  $\xi$  & $\eta$&  $\xi$  & $\eta$ \\ 
  \hline
  \hline\noalign{\smallskip}
CMOS9 & -0.6183 & 0.5186 & -0.6183 & 0.3464 & -0.5004 & 0.3464 & -0.5004 & 0.5186 \\
CMOS10 & -0.474 & 0.5186 & -0.474 & 0.3464 & -0.356 & 0.3464 & -0.356 & 0.5186 \\
CMOS11 & 0.1058 & -0.3925 & 0.1058 & -0.5647 & 0.2237 & -0.5647 & 0.2237 & -0.3925 \\
CMOS12 & 0.2501 & -0.3925 & 0.2501 & -0.5647 & 0.368 & -0.5647 & 0.368 & -0.3925 \\
  \noalign{\smallskip}\hline
  \hline
\end{tabular}
\end{table}

\section{Gaia SQL Query} \label{sect:sql}
\begin{lstlisting}
SELECT  gaia_source.source_id,gaia_source.ra,gaia_source.ra_error,gaia_source.dec,gaia_source.dec_error,gaia_source.pm,gaia_source.pmra,gaia_source.pmra_error,gaia_source.pmdec,gaia_source.pmdec_error,gaia_source.phot_g_mean_mag,gaia_source.phot_g_mean_flux,gaia_source.phot_g_mean_flux_error,gaia_source.ruwe,gaia_source.ecl_lon,gaia_source.ecl_lat,gaia_source.phot_bp_mean_mag,gaia_source.phot_rp_mean_mag,gaia_source.bp_rp
FROM gaiadr3.gaia_source 
WHERE (gaiadr3.gaia_source.parallax<=0.2 AND gaiadr3.gaia_source.parallax_error<=0.2 AND gaiadr3.gaia_source.visibility_periods_used>=8 AND gaiadr3.gaia_source.phot_g_mean_flux_over_error>=50 AND gaiadr3.gaia_source.phot_bp_mean_flux_over_error>=20 AND gaiadr3.gaia_source.phot_rp_mean_flux_over_error>=20 AND gaiadr3.gaia_source.phot_g_mean_mag BETWEEN 8 AND 15 AND gaiadr3.gaia_source.phot_bp_rp_excess_factor>=1.0 + 0.015*POWER(bp_rp,2) AND gaiadr3.gaia_source.phot_bp_rp_excess_factor<=1.3 + 0.06*POWER(bp_rp,2))
\end{lstlisting}

\begin{acknowledgements}
We acknowledge the support by National Key R\&D Program of China (No. 2022YFF0503403, 2022YFF0711500), the support of National Nature Science Foundation of China (Nos. 11988101, 12073047, 12273077, 12022306, 12373048, 12263005), the support from the Ministry of Science and Technology of China (Nos. 2020SKA0110100),  the science research grants from the China Manned Space Project (Nos. CMS-CSST-2021-B01, CMS-CSST-2021-A01), CAS Project for Young Scientists in Basic Research (No. YSBR-062), and the support from K.C.Wong Education Foundation. This work is based on the mock data created by the CSST Simulation Team, which is supported by the CSST scientific data processing and analysis system of the China Manned Space Project.

\end{acknowledgements}

\bibliographystyle{raa}
\bibliography{ref}

\end{document}